\documentclass[twocolumn]{aastex63}

\newcommand{\hi}{{\sc H\,i}}
\newcommand{\fermi}{{\em Fermi}}
\newcommand{\fermilat}{{\em Fermi}-LAT}
\newcommand{\gray}{$\gamma$-ray}
\newcommand{\grays}{$\gamma$-rays}
\newcommand{\GP}{{\sc GALPROP}}

\usepackage[T1]{fontenc}
\usepackage{amsmath}

\received{}
\revised{}
\accepted{}
\submitjournal{ApJ}

\shorttitle{Cosmic Ray Sources in the $\gamma$-Ray Sky}
\shortauthors{Johannesson \& Porter}
\graphicspath{{./}{figures/}}
\begin{document}

\title{Signatures of Recent Cosmic-Ray Acceleration in the High-Latitude $\gamma$-Ray Sky}

\correspondingauthor{Gu{\dh}laugur J{\'o}hannesson}
\email{gudlaugu@hi.is}

\author[0000-0003-1458-7036]{Gu{\dh}laugur J{\'o}hannesson}
\affiliation{Science Institute, University of Iceland, IS-107 Reykjavik, Iceland}
\affiliation{Nordita, KTH Royal Institute of Technology and Stockholm University, Roslagstullsbacken 23, SE-106 91 Stockholm, Sweden}

\author[0000-0002-2621-4440]{Troy A. Porter}
\affiliation{W. W. Hansen Experimental Physics Laboratory and Kavli Institute for Particle Astrophysics and Cosmology, \\ Stanford University, Stanford, CA 94305, USA}

%% Mark off the abstract in the ``abstract'' environment. 
\begin{abstract}

  Cosmic-ray (CR) sources temporarily enhance the relativistic particle density in their vicinity over the background distribution accumulated from the Galaxy-wide past injection activity and propagation.
  If individual sources are close enough to the solar system, their localised enhancements may present as features in the measured spectra of the CRs and in the associated secondary electromagnetic emissions.
  Large scale loop like structures visible in the radio sky are possible signatures of such nearby CR sources.
  If so, these loops may also have counterparts in the high-latitude \gray{} sky.
  Using $\sim$10 years of data from the {\em Fermi} Large Area Telescope, applying Bayesian analysis including Gaussian Processes, we search for extended enhanced emission associated with putative nearby CR sources in the energy range from 1 GeV to 1 TeV for the sky region $|b| > 30^\circ$.
  We carefully control the systematic uncertainty due to imperfect knowledge of the interstellar gas distribution.
  Radio Loop~IV is identified for the first time as a \gray{} emitter and we also find significant emission from Loop~I. 
Strong evidence is found for asymmetric features about the Galactic $l = 0^\circ$ meridian that may be associated with parts of the so-called ``\fermi\ Bubbles'', and
some evidence is also found for \gray{} emission from other radio loops.
  Implications for the CRs producing the features and possible locations of the sources of the emissions are discussed.

\end{abstract}

%% Keywords should appear after the \end{abstract} command. 
%% See the online documentation for the full list of available subject
%% keywords and the rules for their use.

\section{Introduction} 
\label{sec:intro}

After more than 100 years of observations, the precise origin of cosmic rays (CRs) and the nature of their propagation in the Galaxy remains as an outstanding question in astrophysics.
The body of evidence suggests that supernova remnants (SNRs) are the main source class \citep[e.g.,][]{2013Sci...339..807A,2014IJMPD..2330013A}, but pulsar wind nebulae have also been observed to accelerate electrons to very high energies \citep[e.g.,][]{2014IJMPS..2860160A, 2017Sci...358..911A}, making them clear candidates as CR sources as well.
Other sources include shocks in stellar winds of massive stars and jets originating from compact objects like neutron stars, stellar black holes \citep{2013A&ARv..21...64D}, massive black holes at the centres of galaxies \citep{doi:10.1146/annurev-astro-081913-040044}, and gamma-ray bursts \citep{2004APh....21..125W}.
In general, hypothesised sources of CRs have in common that they are usually discrete in both time and space, resulting in highly localised spatio-/temporal CR injection into the interstellar medium (ISM).

Following acceleration at the sources, the CRs scatter off magnetic fields in the Galaxy in a process that can be approximated as diffusion \citep[see, e.g.,][for a review]{GrenierEtAl:2015}.
This diffusive propagation scrambles directionality, hence the CR data do not provide information about the spatial location of the individual sources.
While propagating the CRs interact with the components of the diffuse ISM, producing $\gamma$-rays from nuclei-nuclei interactions and electron-bremsstrahlung emission on the interstellar gas, and electrons inverse Compton (IC) scattering the interstellar radiation field (ISRF).
For energies $\lesssim 30$~TeV, the $\gamma$-rays travel mostly unimpeded from their origin \citep[e.g.,][]{2006ApJ...640L.155M,2018PhRvD..98d1302P} and can provide the crucial directional information for probing the detailed spatial distribution of the CRs.
This has been utilised to infer the CR distribution in the nearby interstellar space \citep[e.g.,][]{LAT-CasCep:2010, Casandjian:2015} and outer Galaxy \citep[e.g.,][]{LAT-3rdquad:2011}, assuming that the CRs vary smoothly with Galactocentric radius.

\citet{PorterEtAl:2019} studied the effect of spatio-/temporal discrete CR sources on predictions for broadband (radio, X-rays, $\gamma$-rays) diffuse emissions.
For nearby discrete CR sources, excess emission over smooth CR source models is expected, which should be most evident for latitudes outside the Galactic plane.
Under the assumption that the source accelerates both nuclei and electrons, the excess is more prominent in the IC emission than the gas-related emission.
Generally, the more rapid energy losses for the high-energy electrons cause bright emissions from the nearby vicinity of the individual sources.
Enhanced emission around discrete sources is also expected from the gas-related component, but with lower intensity.
For both processes, the emission is expected to be spectrally harder than the background, and would present as broadly distributed on the sky, likely with low surface brightness.
Extracting such signals from the GeV photon data is difficult because of the low statistics, low spatial resolution, and {\it a-priori} unknown distribution of the emissions on the sky.
Detections of such features would be a clear signature of recent CR source activity and provide the important evidence for the origin of CRs and their recent history.

This paper describes our analysis of the high-latitude $\gamma$-ray sky observed by the Large Area Telescope (LAT) \citep[][]{AtwoodEtAl:2009} onboard the \fermi\ Gamma-ray Space Telescope.
We use Gaussian Processes \citep[GPs,][]{Cressie:1993} to model the sky distribution of the structured gas and smooth IC components illuminated by the CRs in the ISM.
GPs have been used extensively in geostatistics where they were originally developed and are better known as kriging \citep{Chiles2018}.
They allow for recovery of structures via their connectedness properties without the imposition of specific spatial templates. 
Because of the high degree of dependence between neighbouring regions in the \gray\ sky -- expected to be spatially correlated due to the CR diffusion in the ISM -- the GPs are an appropriate tool for our analysis that is attempting to identify extended structures with {\it a-priori} unknown distributions on the sky.
We identify several features of enhanced flux, including radio Loops I and IV \citep[see, e.g.,][and references therein]{VidalEtAl:2015}, as well as extended features that may be associated with parts of the so-called ``\fermi\ Bubbles'' \citep[FBs;][]{2010ApJ...724.1044S}.%,DoblerEtAl:2010}.
The paper is organised as follows.
The data and methods are described in Sec.~\ref{sec:data_model}, which also contains results from dedicated simulations to test the method.
The results are described in Sec.~\ref{sec:results} and we discuss and summarise in Secs.~\ref{sec:discussion} and~\ref{sec:summary}, respectively.

\section{Data and Model} 
\label{sec:data_model}

\subsection{\fermilat\ data and fitting procedure}
\label{subsec:data}

The LAT is a pair conversion telescope sensitive to photons in the energy range from around 30~MeV to greater than few hundred GeV \citep{AtwoodEtAl:2009}.
Due to its surveying observing strategy and wide field of view, the LAT data have nearly uniform coverage of the entire sky and are well suited for studies of large scale structures.
This study uses $\sim$10 years of the most up-to-date data release of P8R3 photon events \citep{BruelEtAl:2018, AtwoodEtAl:2013}.
The time range is identical to that used in the second data release \citep[DR2;][]{4FGL-DR2} of the \fermilat\ fourth source catalog \citep[4FGL;][]{4FGL-paper}.
This ensures that the source properties can be used directly without re-adjusting for possible variation caused by known time evolution of the most common source class, the active galactic nuclei (AGN).
This analysis uses the %{\tt SOURCE}
{\tt ULTRACLEANVETO} photon event type and the associated
%{\tt P8R3\_SOURCE\_V2}
{\tt P8R3\_ULTRACLEANVETO\_V2} instrument response functions (IRFs).  

The data are binned into a HEALPix\footnote{\url{https://healpix.jpl.nasa.gov/}} grid with $N_{\mathrm side}$ of 256 (order~8), having a spatial resolution of $\sim$$0.25^\circ$.
The data are analysed independently in 4 coarse energy bins (1--3~GeV, 3--10~GeV, 10--60~GeV, and 60--1000~GeV) with each split further into 4 equally spaced logarithmic sub-bins.
This energy binning is a compromise between a fine energy resolution and having enough statistics for each bin to constrain the spatial distribution of the model.
The sub-bins allow for spectral separation of the gas/structured and smooth components of the model, which are expected to have different spectral shapes (see Sec.~\ref{subsec:model} for the model description).
To avoid complications arising from modelling the Galactic plane, the analysis is restricted to latitudes $|b| > 30^\circ$.
This latitude cut is motivated by the results of \cite{PorterEtAl:2019} that showed the effect of discrete CR sources to be largest away from the plane.  
It also removes most of the contribution of molecular gas and the Galaxy in general, and allows for a simpler model of the sky.

The analysis is Bayesian in nature and parameter sampling and optimisation is performed using the Stan platform for statistical modelling\footnote{https://mc-stan.org/}.
The no U-turn sampler (NUTS) provides the required efficiency to accurately sample the high number of model parameters in a reasonable time.
For each run, 30 chains are started containing 400 samples each for a total of 12000 points per analysis.
For performance reasons, the chains are all started from the same warm-up phase that also contains 400 points as explained in the manual for the Python interface of Stan.
Independence is achieved by starting each chain with a unique seed and adding an extra warmup phase of 25 points.
Random spot checks of chains show that they are fully developed using this method, and the $\hat{R}$-statistics calculated by Stan return 1 for all variables over the runs indicating convergence.

\subsection{Model}
\label{subsec:model}

The model is based on the simple assumption that the $\gamma$-ray emission from the high-latitude sky can be separated into distinct components: (i) emission from point-like and slightly extended sources as listed in the 4FGL-DR2, (ii) structured emissions originating from interactions between CRs and the interstellar gas, and (iii) a smoothly distributed emissions from interactions between CRs and the ISRF, as well as from unresolved sources and irreducible background in the data.
This is well motivated by theoretical expectations as well as previous studies of the LAT data \citep[e.g.,][]{DiffusePaperII:2012, AckermannEtAl:2012}.  
For the first component, the source properties are kept fixed to that listed in the 4FGL-DR2.
Because most of the resolved point sources in the region of interest are AGN, and this source class is known for nearly all cases to have curved spectra when statistics are high enough \citep{4FGL-paper}, sources that are listed as having a preferred power-law spectra are modelled using the provided log-parabola spectrum instead, so as to not bias the analysis by ignoring curvature of faint sources.
This criterion was also applied when deriving the interstellar emission model used for the 4FGL\footnote{\url{https://fermi.gsfc.nasa.gov/ssc/data/analysis/software/aux/4fgl/Galactic_Diffuse_Emission_Model_for_the_4FGL_Catalog_Analysis.pdf}}.

We model the gas column density distribution using four different tracers: the HI4PI 21-cm \hi{} line emission \citep{HI4PI-paper}, the dust optical depth at 353~GHz ($\tau_{353}$), dust radiance \citep{Planck-dust:2016}, and optical extinction $A_V$ \citep{Planck-dust-dl:2016}.
The \hi{} line emission should be the most accurate in determining the neutral hydrogen column density for an optically thin medium, but it will miss any molecular contribution, and correcting for optical thickness effects is not trivial.
Dust emission/extinction have been shown to be better tracers of the total gas column density \citep[e.g.,][]{GrenierEtAl:2005,LAT-CasCep:2010,LAT-3rdquad:2011}, but varying and uncertain dust properties make them susceptible to other systematic errors \citep{RemyEtAl:2017,RemyEtAl:2018}.
For this work, we use multiple templates to reduce the effects of the different systematics.
%To reduce the effect of these different systematics, for this study we use a varied set of templates.

For each of the four gas tracers, template maps are generated under different assumptions for converting the respective data to gas column densities.
The \hi{} data are converted to column densities under the assumption of a uniform spin temperature of $T_S=100$~K and $T_S=50$~K \citep[see, e.g.,][for more details]{DiffusePaperII:2012}, respectivley, producing two slightly different templates.
Preliminary investigation showed that the difference between templates using the optically thin assumption and $T_S=100$~K were minor in our regions of interest, hence the former is not used in this analysis.
For the dust emission, $D$, the gas column density, $G$, is calculated using
\begin{equation}
    G = X_d \left( D - d_0 \right)^{1/\alpha}
    \label{eq:dustToGas}
\end{equation}
where $X_d$ is the dust-to-gas ratio, $d_0$ is a global offset, and $\alpha$ accounts for possible non-linearity in the conversion from gas to dust.
This form has been used before in analysis of $\gamma$-ray data where the best fit value was $\alpha\sim$1.4 \citep{HayashiEtAl:2019}.
For this analysis, the value of $\alpha$ is chosen to be one of 1.0, 1.2, and 1.4 and the values of $X_d$ and $d_0$ are determined using a maximum likelihood fit. 
For technical reasons, the parameters are determined in the fit using the inverse of Eq.~\ref{eq:dustToGas}.
  That is, the dust column density is the data and $G$ is the optically thin \hi{} column density.
We use the Student's t likelihood to reduce the effect of outliers caused by missing molecular gas with $\nu=10$ degrees of freedom.
We use the uncertainties reported for the various dust tracers when forming the likelihood, except for the $A_V$ data.
%The dust data
%are all, except the $A_V$ data, supplemented with the uncertainties reported for the 
%an uncertainty scaled in units of in forms of standard deviation which we use in the fits.
A per-pixel uncertainty for this tracer is not provided.
Therefore, we assume for $A_V$ a fixed 3\% uncertainty, which is the mean for the $\tau_{353}$ data.
The fit is limited to $|b|>30^\circ$ and, to further reduce the effects of the missing molecular gas, the fit is iterated several times removing pixels where the model underestimates the observed dust data by $>$5 standard deviations in each iteration.
The resulting parameters are given in Table~\ref{tab:dustPars}.
Because the model, described below, includes an overall normalisation of the gas component, the exact value of $X_d$ has no effect on the final estimated \gray{} intensity and only influences the absolute normalisation of the emissivity spectrum.
By using the optically thin \hi{} data in the fit, we minimise this unavoidable effect.

To facilitate the discussion, each map is labelled using the identifier given in the table appended with $10\alpha$, e.g., Rad12 is the map based on radiance data with $\alpha=1.2$.
For the \hi{} data, the labels are HI50 and HI100 for a spin temperature correction of 50~K and 100~K, respectively.
The heavier elements in the ISM are assumed to be well mixed with the hydrogen gas and are included to give the total gas column densities utilised for the data analysis.
To do this we multiply each template by a factor $1.42$, assuming a hydrogen mass fraction of 70\%.

\begin{table}
    \caption{Parameters used for the determination of the different gas templates based on dust based observations.}
    \label{tab:dustPars}
    \centering
    \begin{tabular}{l|c|c|c}
        Dust map & $\alpha$ & $X_d$\tablenotemark{a} & $d_0$\tablenotemark{b} \\
        \hline
        $\tau_{353}$\tablenotemark{c} & $1.0$ & $1.14 \times 10^{26}$ & $-2.6 \times 10^{-7}$ \\
         Tau\tablenotemark{f} & $1.2$ & $1.64 \times 10^{25}$ & $1.6 \times 10^{-7}$ \\
          & $1.4$ & $4.17 \times 10^{24}$ & $4.6 \times 10^{-7}$ \\
          \hline
        Radiance\tablenotemark{d} & $1.0$ & $4.8 \times 10^{27}$ & $4.4 \times 10^{-11}$ \\
         Rad\tablenotemark{f} & $1.2$ & $2.9 \times 10^{26}$ & $4.6 \times 10^{-9}$ \\
          & $1.4$ & $4.8 \times 10^{25}$ & $1.3 \times 10^{-8}$ \\
        \hline
        $A_V$\tablenotemark{e} & $1.0$ & $8.6 \times 10^{20}$ & $-0.095$ \\
         Av\tablenotemark{f} & $1.2$ & $9.5 \times 10^{20}$ cm$^2$ & $-0.011$ \\
          & $1.4$ & $1.1 \times 10^{19}$ cm$^2$ & $0.062$ \\
    \end{tabular}
    \tablenotetext{a}{In appropriate units so the output column density is cm$^{-2}$.}
    \tablenotetext{b}{In units of the dust map, see below.}
    \tablenotetext{c}{COM\_CompMap\_Dust-GNILC-Model-Opacity\_2048\_R2.01.fits, no units.}
    \tablenotetext{d}{COM\_CompMap\_Dust-GNILC-Radiance\_2048\_R2.00.fits, in units of W~m$^{-2}$~sr$^{-1}$.}
    \tablenotetext{e}{COM\_CompMap\_Dust-DL07-AvMaps\_2048\_R2.00.fits, in units of magnitudes.}
    \tablenotetext{f}{Identifier for the different maps}
\end{table}

To account and search for possible non-uniformity in the density of CRs, the directional emissivity\footnote{Defined as the intensity of the gas component divided by the gas column density.} of the gas component and the intensity of the smooth component are modelled using GPs.
GPs are spatial random processes that build upon the Gaussian distribution and are constrained using a covariance function that depends in most cases on the distance between two spatial points.
The spatial binning of the sky for the GPs are based on the HEALPix pixelisation, using an $N_{\mathrm side}$ parameter of 4 and 8 for the  directional gas emissivity and intensity of the smooth component, respectively.
Finer binning is used for the smooth component because it is expected to vary more rapidly due to the fast cooling of electrons at the highest energies.
The resolution, and hence the characteristic size of the GPs, is deliberately kept larger than the size of the LAT point-spread-function (PSF).
For the current data selection, the 68\% containment radius of the PSF is $\sim$$1^\circ$ at 1~GeV and reduces for higher energies\footnote{\url{https://www.slac.stanford.edu/exp/glast/groups/canda/lat_Performance.htm}}.
This is considerably smaller than the characteristic radius of the pixels in the GPs, which are $\sim$$7^\circ$ and $\sim$$15^\circ$ for the smooth intensity and gas emissivity, respectively.
To avoid sharp edges at the pixel boundaries, the emissivity and intensity values are interpolated to a HEALPix grid with the same resolution as the data ($N_{\mathrm side}=256$, see Sec.~\ref{subsec:data}) using the HEALPix interpolation facility.

As an implementation convenience, we use the GPs to model the directional emissivity and smooth intensity as correction factor spatial distributions that are applied to otherwise uniform emissivity and intensity power-law spectra.
In other words, the GPs model the spatially dependent departures from isotropic sky distributions for the emissivity and smooth intensity components.
We take for the emissivity a spectrum with power-law index $-2.7$ that is normalised to $3.5\times10^{-33}$~MeV$^{-1}$~s$^{-1}$~sr$^{-1}$ at 10~GeV. 
This is based on the emissivity derived from high-latitude LAT data \citep{Casandjian:2015}.
For the smooth intensity, the power-law index is $-2.2$ and it is normalised to $9\times10^{-8}$~MeV$^{-1}$~cm$^{-2}$~s$^{-1}$~sr$^{-1}$ at 100~MeV, following the derived isotropic background spectrum\footnote{\url{https://fermi.gsfc.nasa.gov/ssc/data/access/lat/BackgroundModels.html}}.
Correspondingly, the model intensity for energy $E$ and direction $\vec{\theta}$ is
\begin{equation}
\begin{split}
    M(E,\vec{\theta}) &= N_g(\vec{\theta}) \left(\frac{E}{E_0}\right)^{\gamma_g} T_g(E,\vec{\theta})\\ 
    &+ N_s(\vec{\theta}) \left(\frac{E}{E_0}\right)^{\gamma_s} T_i(E) + T_f(E,\vec{\theta}),
\end{split}
\label{eq:model}
\end{equation}
where $N_{g,s}$ are the spatial correction factors for the structured (gas) and smooth templates, respectively, that are based on the GPs and described further below.
$T_g$ is the gas template map multiplied with the power-law uniform emissivity spectra, $T_i$ is the isotropic template for the smooth component, and $T_f$ is a fixed template containing the sources in the 4FGL-DR2.
Both the gas and the smooth template are additionally scaled with a power-law correction, $\gamma_{g,s}$, to globally adjust their spectral shape for the two hemispheres and within each of the larger energy bins (see Sec.~\ref{subsec:data}), with $E_0$ fixed to the geometrical mean energy of the coarse bin being analysed.
The input spectrum chosen above is thus allowed to be adjusted in the analysis, freely between coarse bins and using the power-law adjustment within coarse bins.
In this analysis, $E$ and $\vec{\theta}$ are discretised and the model is evaluated in the same binning as the data, which is described in Sec.~\ref{subsec:data}.  
The model intensity is evaluated at the centre of each pixel and geometric mean value of each of the finer sub-energy bins.

To avoid negative values in the model, $N_g$ and $N_s$ are modelled as the exponential of the GPs
\begin{equation}
    N_x = f_{\mathrm{int}}\left( \mu_x e^{\tilde{N}_x}\right),
\end{equation}
where
\begin{equation}
    \tilde{N}_x \sim \mathcal{N}(0, \mathbf{C}_x)
\end{equation}
is the actual GP.
Here the subscript $x$ stands for either $g$ or $s$,
$f_{\mathrm{int}}$ is the interpolation function scaling the GP to the full resolution of the data, and
  $\mu_x$ is the exponential of the mean value of the respective GP.
  Modelling the mean is necessary to ensure that the fluctuations are properly centred.
  This formulation is similar to, but not exactly the same as, a log-normal process.
  The elements of the covariance matrix $\mathbf{C}_x$ are modelled with
  a Mat{\'e}rn \citep{Matern:1986} function
\begin{equation}
    C_{x_{i,j}} = 0.1 \beta_x^2 e^{-\frac{d_{i,j}}{2\rho_x}}
    \label{eq:covariance}
\end{equation}
where $d_{i,j}$ is the angular distance between pixels $i$ and $j$.
The characteristic distance $\rho_x$ allows for flexibility in the correlation length, while the amplitude parameter $\beta_x$ sets the magnitude of the fluctuations.

In addition to the free parameters of the GPs that are constrained by their respective multi-normal distributions, the model has 8 more parameters, $\rho_x$, $\beta_x$, $\mu_x$, and $\gamma_x$ (the subscript $x$ stands for either $g$ or $s$) that require a prior.
The inverse gamma-function is used for $\rho_x$ with the shape and scale parameters both set to 5, with the distances in the covariance function ($d_{i,j}$ in Eq.~\ref{eq:covariance}) scaled such that the peak of the distribution corresponds to $50^\circ$ for $\rho_g$ and $15^\circ$ for $\rho_s$\footnote{The inverse gamma-function with shape and scale of 5 peaks at around 1 with a range of a factor of few. Rather than adjusting the gamma-function parameters, it is more convenient to adjust the scale of the distance to have the prior appropriate for the binning. We picked $50^\circ$ and $15^\circ$ based on the size of the pixels used for the respective GPs.  The results are insensitive to the exact value of these parameters: a factor of 2 up or down will not alter them significantly.}.
The prior for $\beta_x$ is the half standard normal distribution and for $\mu_x$ it is a log-normal distribution based on the standard normal distribution.
Finally, the prior for $\gamma_x$ is a normal distribution with a mean 0 and standard deviation of 0.1.
The half standard normal prior for $\beta_x$ ensures that the fluctuations are suppressed to reduce the effects of random statistical variation in the data.

For the likelihood calculations, the model $M(E,\vec{\theta})$ has to be convolved with the IRFs, which is a computationally intensive operation.
To save time, only the model templates $T_g$, $T_s$, and $T_f$ are convolved with the IRFs, and only once at the start of the analysis.
This optimisation assumes that  modifications to the model-predicted counts are directly proportional to changes in the intensities.
The validity of this approach has been tested with simulated data using a best-fit output model from the analysis procedure as input templates ($T_g$ and $T_s$ in Eq.~\ref{eq:model}), and re-running the analysis (see Appendix~\ref{app:simulations} for details).
If convolution with the IRFs were to produce strong effects, this new run would show significant variations from the best-fit results.
However, the resulting GPs were consistent with the identity within the statistical uncertainties.
The small changes are dominated by overall normalisation and spectral modifications with hardly any spatial variations.  
These small changes are nearly eliminated by modifying the input spectrum of the templates $T_g$ and $T_s$ with the mean spectrum of $N_g$ and $N_s$, respectively, in each coarse energy bin and hemisphere.
Our results on the spatial distributions with these simplifying assumptions are therefore robust, and there is no need to do the IRF convolution each time the likelihood is evaluated\footnote{This is because the pixel size of the GPs are much larger than the size of the PSF.  Using finer resolution GPs would require an IRF convolution for each likelihood evaluation.}.
In this work, the model templates $T_g$, $T_s$, and $T_f$ are converted from intensities to counts using the GaRDiAn code \citep{DiffusePaperII:2012}.
It takes into account the energy and spatially dependent exposure and PSF of the LAT, and accounts for the energy dispersion of photons.

\subsection{Simulations} 
\label{subsec:simulations}

We employ simulations to validate the modelling assumptions and analysis pipeline.
The simulations test the effectiveness of the spectral corrections to adjust the model to the data, the separation of components, the sensitivity of the GPs to extracting enhanced emissions from recent discrete CR source injection, and the effect of the fixed point source component.
Only a subset of the simulation results are presented here -- enough to demonstrate the sensitivity and component separation -- with the remainder given in Appendix~\ref{app:simulations}.
To summarise, the simulations demonstrate that the spectral correction is effective and a single iteration applying the mean spectrum to the input template is enough to get a reliable result.
The simulations also demonstrate that reasonable variations in properties of the sources in 4FGL-DR2 do not affect the results of the GPs.

The simulations are made at the pixel level.
The GaRDiAn code is used to convolve the input intensity maps into expected counts binned in both sky position and energy.  The simulated data are then drawn from Poisson distributions whose rates are given with the binned expected counts.
 This method is much simpler than using the {\it Fermi} gtobssim tool\footnote{https://raw.githubusercontent.com/fermi-lat/fermitools-fhelp/master/gtobssim.txt}, which simulates each photon individually, and is sufficient for this work where the photon data are analysed in a binned fashion.

\begin{table}
    \centering
    \caption{The parameters of the three enhanced emission regions used in the simulations.}
    \begin{tabular}{c|c|c}
    Centre $(l,b)$ [deg] & Width [deg] & Magnitude\tablenotemark{a} \\
    \hline
    $(20,70)$ & 15 & 20\%  \\
    $(240,-50)$ & 7 & 30\%  \\
    $(330,-30)$ & 20 & 15\%  \\
    \end{tabular}
    \tablenotetext{a}{The fraction for the smooth component at 10~GeV}
    \label{tab:BumpParameters}
\end{table}

While our method is Bayesian and requires the evaluation of the entire posterior distribution for the data, we use a simplified analysis for the simulations finding only the maximum of the posterior distribution.
   This is much faster and enables easy analysis of many simulations to test for biases and estimate the statistical uncertainty through a frequentist approach.  We make 2000 simulations for each case and the maximum posterior value for each parameter is stored.
 The resulting distribution for each parameter is then analysed to estimate the statistical uncertainty and look for biases in the results by comparing to the truth.

\begin{figure*}
    \centering
    \includegraphics{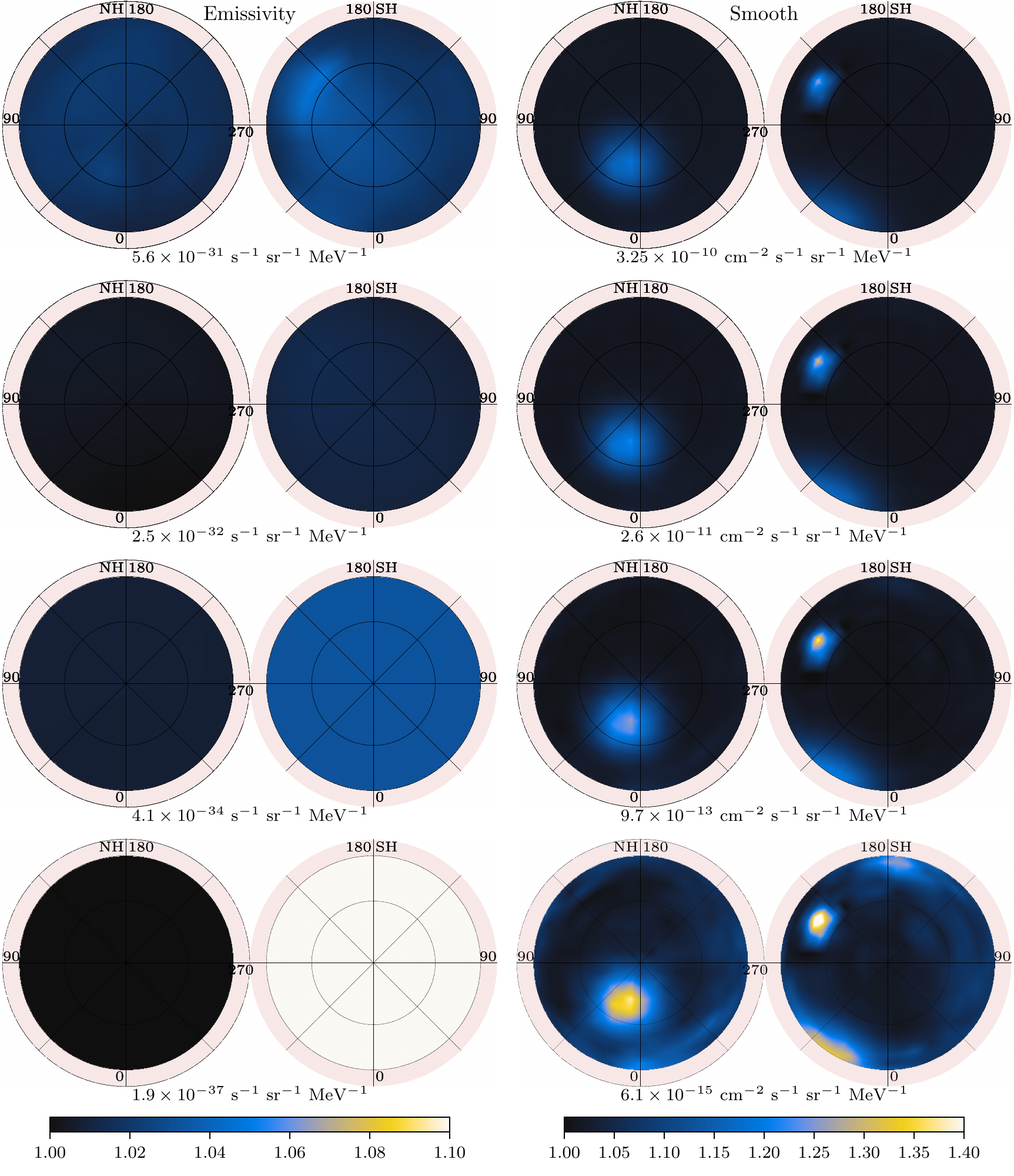}
    \caption{The results from analysis of 2000 simulations based on an idealistic model with three regions of enhanced emission using the same column density map for the simulation and analysis.
      Shown is the emissivity of the gas component (left) and the intensity for the smooth component (right) at the geometric mean energies of the four energy bins (1.7~GeV, 5.5~GeV, 24~GeV, and 240~GeV from top to bottom).
      The maps are in Galactic coordinates and use the orthographic projection with the north hemisphere (NH) on the left and the south hemisphere (SH) on the right.
      The graticules are spaced $30^\circ$ apart in latitude and $45^\circ$ in longitude.
      The fractional colour scale for each column is shown at the bottom and the units are shown below each panel.}
    \label{fig:Bumps_gas}
\end{figure*}

To test the ability of our technique to extract an expected signal of enhanced emissivity of the gas and intensity of the smooth background, we first make an idealised simulation/analysis.
The background consists of emission from a gas component based on the HI100 template with uniform emissivity power-law spectra with index of $-2.7$ and an isotropic component whose intensity spectra is a power-law with an index of $-2.2$. 
Three localised enhanced emission contributions are modelled using Gaussian spatial distributions with varying width, central location, and magnitude.
These regions are added to both the structured and smooth components as a fraction of the underlying background.

We make the spectra of these regions harder than the background components, which is similar to what we might expect if there are such signatures in the LAT data.
For the smooth component the enhanced emission regions have indices $-2.0$ (0.2 dex harder) while for the gas component, the indices are set to $-2.6$ (0.1 dex harder).
The intensity/emissivity of each region is set as a fraction of the underlying background, and the fraction for the gas emissivity is always a third of that of the smooth component.
The parameters for each region are listed in Table~\ref{tab:BumpParameters}.

Figure~\ref{fig:Bumps_gas} shows the results from our analysis of the 2000 simulations using this model for our northern hemisphere (NH) and southern hemisphere (SH) regions of interest.
Shown are both the smooth intensity as well as the gas emissivity.
The method is clearly able to pick up even the faintest of the three enhanced regions in the smooth component at approximately the correct level.
There are some deviations between the input truth and the model, because the spatial resolution of the GPs is not as fine as that of the mock data and hence does not precisely trace the peaks and exact shapes.
But, the overall agreement within the limitations is very good.
The same can not be said for the enhancements in the gas emissivity which are not as cleanly recovered.
This is due to a combination of factors, including limited statistics, the limited spatial resolution in the model for the structured component, and that the enhanced regions are less distinct from the background emissions.
The separation of the gas and smooth components is difficult at the highest energies due to limited statistics.
This is a general feature for all the simulations and is caused by the limited collecting area of the LAT in the highest-energy bin that we employ for this analysis. 
Because of the prior on $\beta_x$, and that the smooth emission is generally more intense in the highest energy bin, features in the gas component are often erroneously assigned to the smooth component.

\begin{figure*}
    \centering
    \includegraphics{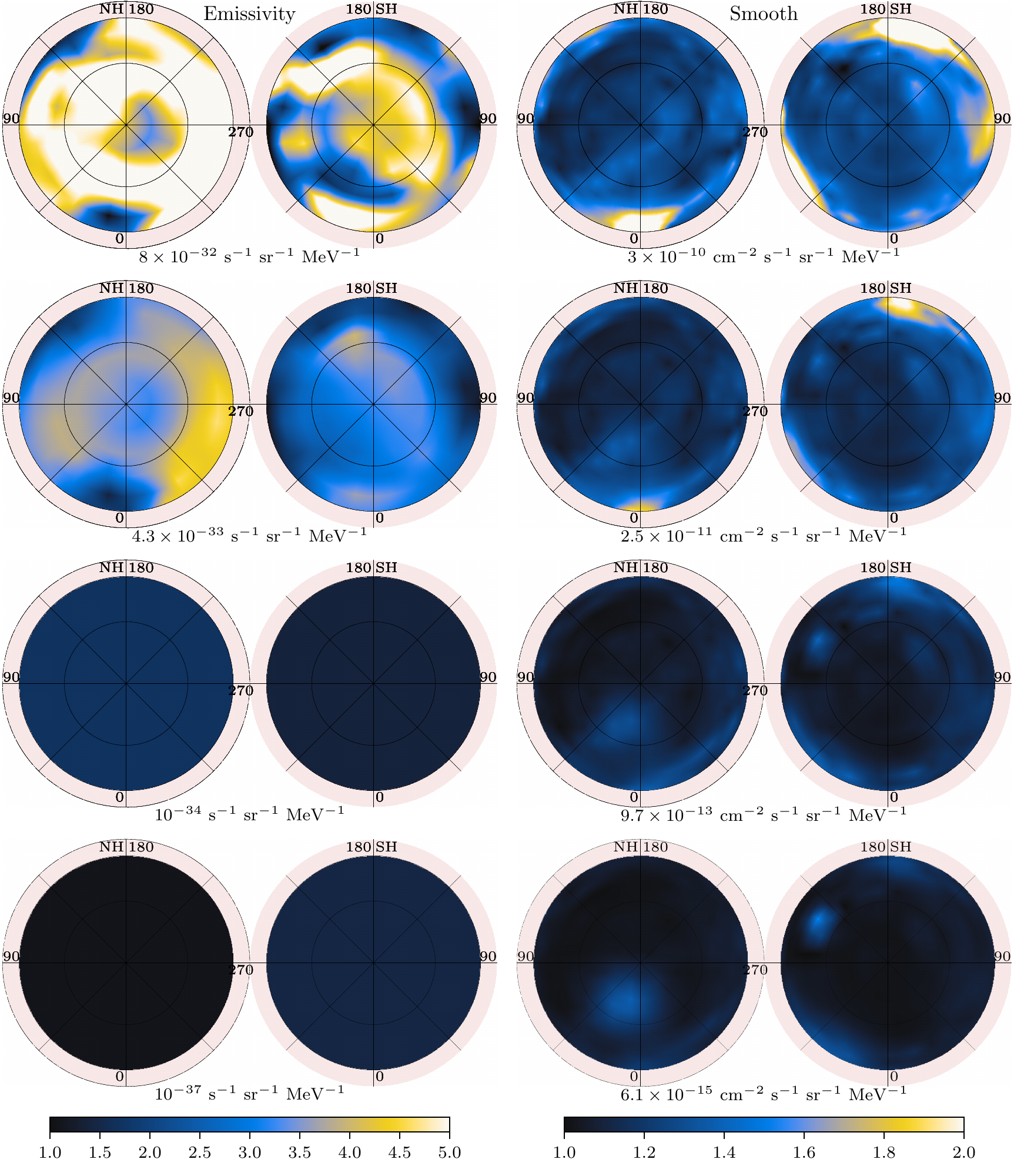}
    \caption{The results from analysis of the 2000 simulations based on the idealised model with three regions of enhanced emission.
      The simulations use the 21-cm HI4PI data, while the analysis on the mock data use the $\tau_{353}$ dust map with $\alpha=1.0$.
      Shown is the emissivity of the gas component (left) and the intensity for the smooth component (right) at the geometric mean energies of the four energy bins (1.7~GeV, 5.5~GeV, 24~GeV, and 240~GeV from top to bottom).
      The maps are in Galactic coordinates and use the orthographic projection with the NH on the left and the SH on the right.
      The graticules are spaced $30^\circ$ apart in latitude and $45^\circ$ in longitude.
      The colour scale differs from that used in Fig.~\ref{fig:Bumps_gas}.
      The results for the highest energy bin are very similar using either representation for the gas component analysing the simulated data.}
    \label{fig:Bumps_dust}
\end{figure*}

The analysis above utilises the same gas column density map that was employed to create the simulated data.
This is not a realistic situation, because the true column density of gas in the Galaxy is imperfectly known.
To examine the systematic effect of this for our method, we take the simulations of the three enhanced emission regions that were created using the HI100 map, and re-run the analysis method using the Tau10 map.
The results are shown in Fig.~\ref{fig:Bumps_dust}.
It is clear that the incorrect gas template has a significant effect on the results.
There are significant fluctuations in both the recovered intensity of the smooth component and the gas emissivity at a level much higher than the injected regions.
Only for the highest two energy bins, where the structured component is negligible for the total emission, do we recover the input.

Because the results are very sensitive to the actual structure of the gas map, it is important to use an appropriate measure to identify sky regions where the analysis performs sub-optimally.
The likelihood value only gives the best performing model, which is unhelpful if none of the models represent the truth.
Visual inspection of residual maps from the simulations does not reveal any apparent structure and therefore they cannot be used to indicate a problem with the fit.
However, analysis of the simulations does show that where the recovery of the truth is problematic, there is a clear anti-correlation between the mean normalisation of the two templates.
Such behaviour is not expected in reality.
True physical changes in the templates should be correlated for combined CR nuclei/lepton injection when the target density is high enough, or independent for the cases of CR nuclei-/lepton-only injection or when one of the target densities (gas/ISRF) is too low for significant \gray{}-emission.

Regular correlation tests, such as the Pearson-r test, applied to the posterior distribution of the parameters are unfortunately not suitable, because there is an anti-correlation between the normalisations of the structured and smooth components even for the idealised case. 
The deviation of the mean normalisation of each pixel, $\bar{N}_{x,j}$, from the mean normalisation of the entire map, $\bar{N}_x$, is the appropriate indicator for our purpose.
This needs also to be scaled by the standard deviation of the normalisation of each pixel.
The final correlation statistic is thus
\begin{equation}
    \kappa_j = \frac{(\bar{N}_{g,j} - \bar{N}_g) (\bar{N}_{s,j} - \bar{N}_s)}{S(N_{g,j}) S(N_{s,j})},
    \label{eq:gas_smooth_correlator}
\end{equation}
where $S(N_{x,j})$ is the standard deviation of the posterior sample for $N_{x,j}$ and $\bar{N}_{x,j}$ is its mean.
This correlation differs from standard image based one that is usually normalised with the standard deviation of the pixels in the image.
Here the normalisation is based on the statistical distribution in each pixel. 
%and measures the correlation in terms of the statistical significance of the means.
The absolute values of the correlations thus provide information about the magnitude of the correlations in terms of statistical significances.

\begin{figure}
    \centering
    \includegraphics{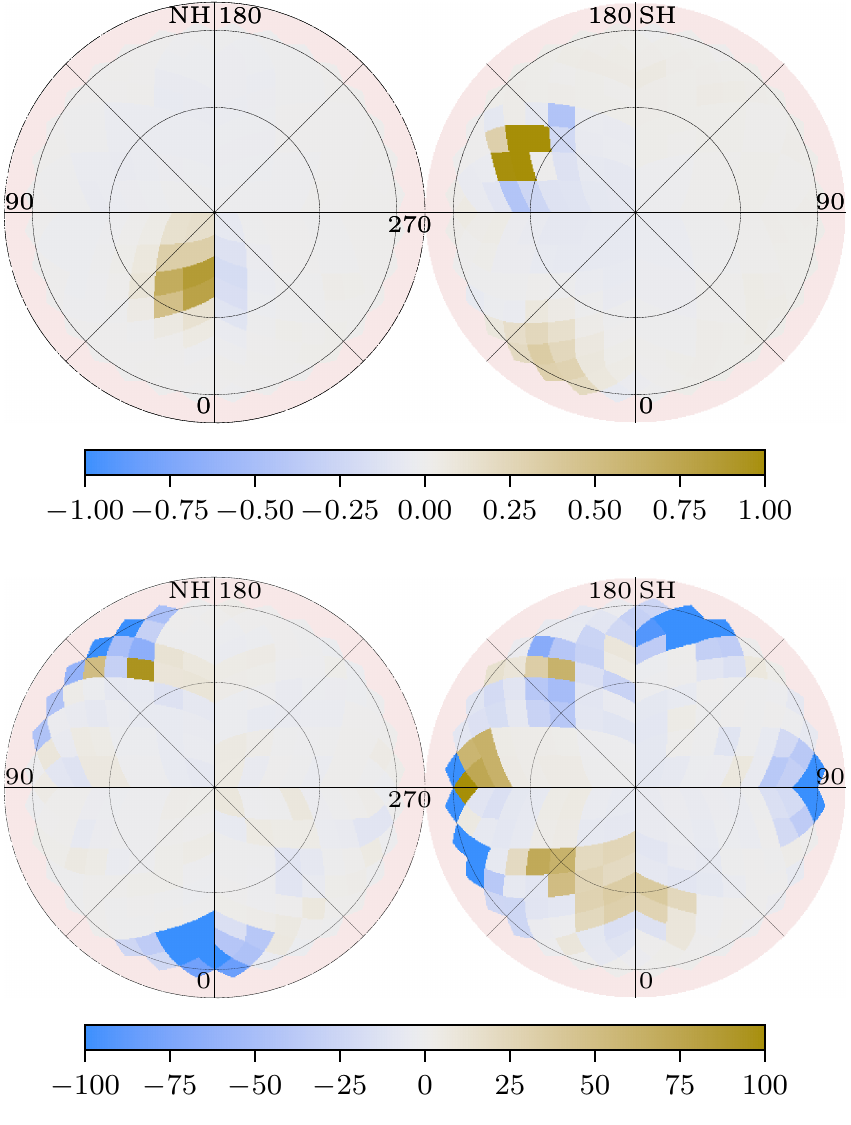}
    \caption{Correlation between the variation from the mean of the emissivity and intensity according to Eq.~(\ref{eq:gas_smooth_correlator}) for the three region simulation.
      Top panel shows the correlations for the lowest energy range for the analysis using the HI100 map, and the bottom panel the same for the analysis using the Tau10 map.
      The correlations are normalised to the statistical error of the analysis, hence the different ranges for the colour scale.
      The maps are in Galactic coordinates and use the orthographic projection with the NH on the left and the SH on the right.
      The graticules are spaced $30^\circ$ apart in latitude and $45^\circ$ in longitude.
    }
    \label{fig:Bumps_correlation}
\end{figure}

Figure~\ref{fig:Bumps_correlation} shows the correlation statistic (Eq.~\ref{eq:gas_smooth_correlator}) for the simulations with the 3 enhanced emission regions, both for the analysis using the HI100 map and the Tau10 map, respectively.
The correlations are clearly positive where the enhanced emissions regions are located using the correct gas column density map as a template, while there are clear negative correlations when employing the incorrect map.
The negative correlations are also much larger in magnitude and align very well with artifacts that are visible in the smooth intensity map in Fig.~\ref{fig:Bumps_dust}.
However, the correlation map is not a perfect indicator.
There are large positive correlations that are associated with regions where both maps are negative.
These are though fewer, and also of smaller absolute magnitude, than the negative ones.
The correlation statistic (Eq.~\ref{eq:gas_smooth_correlator}), in combination with the maximum likelihood of the models, can therefore be used to filter out regions where the results are susceptible to systematic error caused by imperfect knowledge of the gas column density.
It is also necessary to interpret the results in terms of all energy bins, because the effects of an incorrect gas template is expected to be smaller at the highest energy bins due to the steeper spectrum of the gas component.

\subsection{Adjusting the gas templates}
\label{subsec:adjust_gas}

\begin{figure*}
    \centering
    \includegraphics{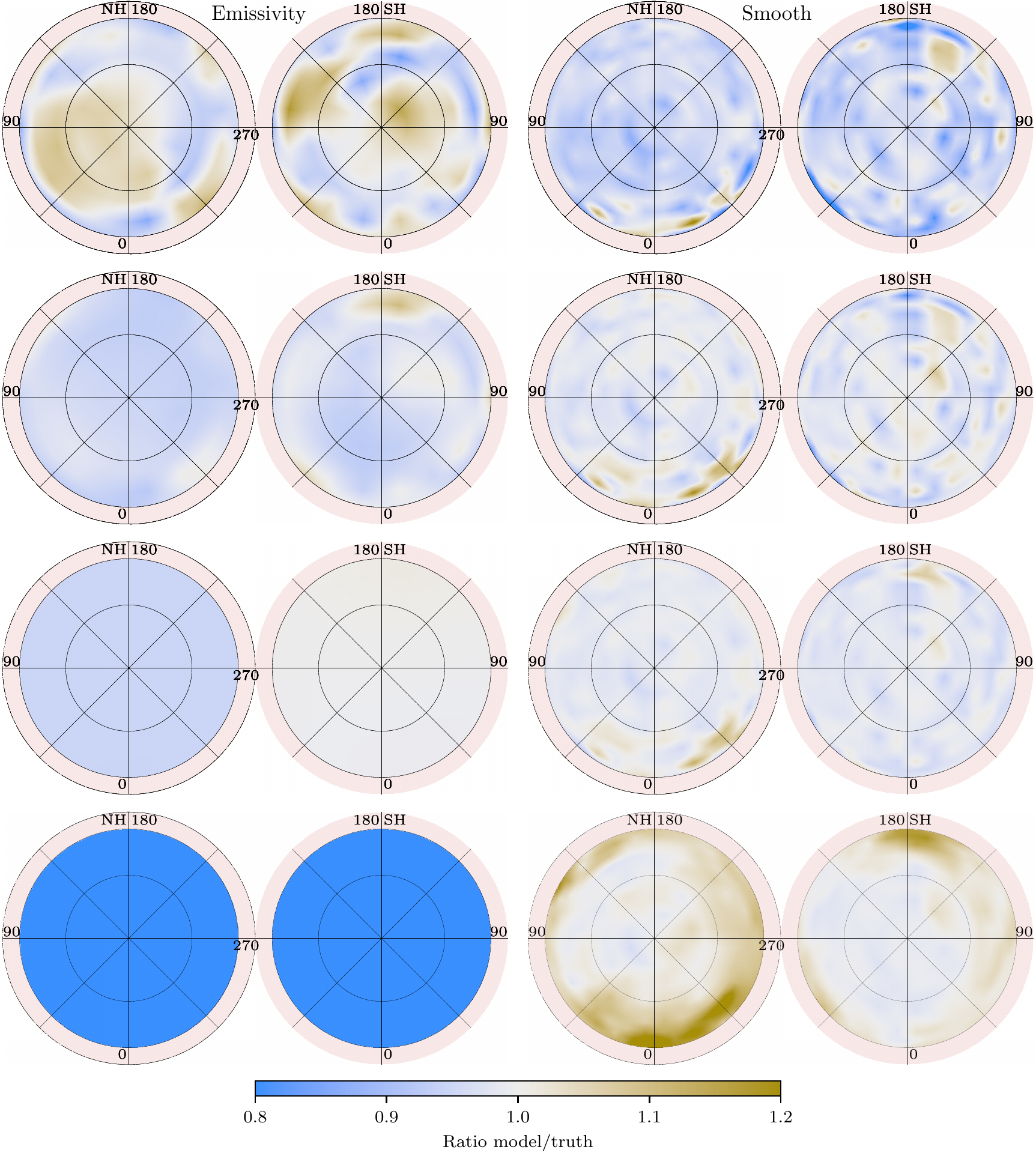}
    \caption{Ratio of the mean of the results when analysing the 2000 simulation of the random gas map using the best-fit likelihood to estimate the correct gas map over the simulation truth.
      The maps are in Galactic coordinates and use the orthographic projection with the NH on the left and the SH on the right.
      The left column shows the ratio of the gas emissivity while the right shows the ratio of the intensity of the smooth component.
      The ratio is calculated at the geometric mean of each of the energy bins, from top to bottom: 1.7~GeV, 5.5~GeV, 24~GeV, 240~GeV.
      The graticules are spaced $30^\circ$ apart in latitude and $45^\circ$ in longitude.
    }
    \label{fig:best_fit_gas_simulation}
\end{figure*}

The individual gas templates listed in Sec.~\ref{subsec:model} are not expected to give a perfect representation of the structured emissions in the \fermilat\ data.
However, it has been shown that a linear combination of these maps can provide a good description of the gas-related $\gamma$-ray emission for small regions of the sky \citep[e.g.,][]{LAT-CasCep:2010,LAT-3rdquad:2011}.
It is also known that the likelihood is a good indicator of the best-fit gas maps.
Therefore, for the LAT data analysis we will use a composite gas column density map created by segmenting the sky into a HEALPix grid with $N_{\mathrm{side}}$ of 8 and selecting in each pixel the gas template whose model is best according to the Akaike information criterion \citep[AIC,][]{Akaike:1974} for that region.
This will produce an ``index'' map into the gas templates.
To avoid sharp boundaries, this coarse grid of ``indices'' is interpolated to the original resolution of the templates, resulting in a linear combination of the maps for all directions, except those directly underneath the centre of pixels in the ``index'' map.

\begin{figure}
    \centering
    \includegraphics{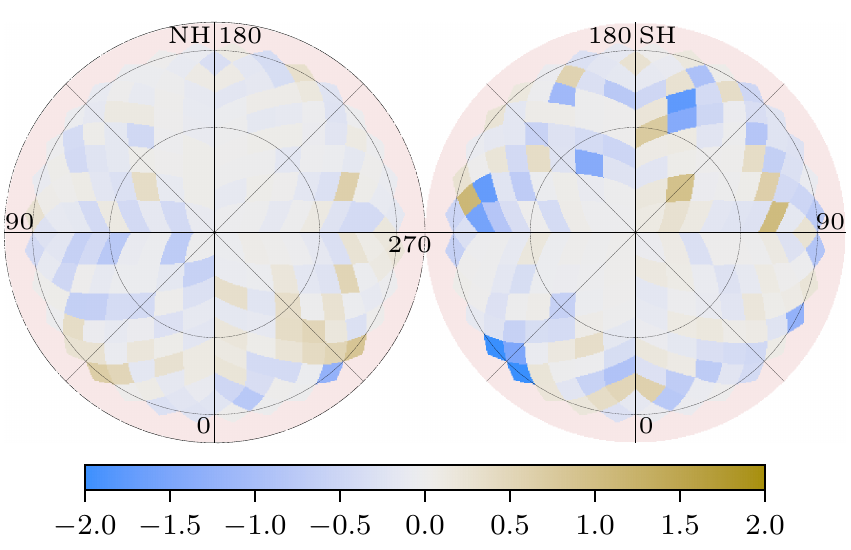}
    \caption{Correlation between the variation from the mean of the emissivity and intensity according to Eq.~(\ref{eq:gas_smooth_correlator}) for the lowest energy bin in the random gas map simulation.
      The maps are in Galactic coordinates and use the orthographic projection with the NH on the left and the SH on the right.
      The graticules are spaced $30^\circ$ apart in latitude and $45^\circ$ in longitude.
    }
    \label{fig:best_fit_gas_correlation}
\end{figure}

We employ simulations to test this composite gas map with the analysis methodology.
We used as input a gas map constructed by selecting an arbitrary ``index'' map.
The simulation is performed using uniform smooth intensity and emissivity.
%, with any deviation from uniformity for the recovered distributions caused by biases in the method.
As before, 2000 simulations are created and each simulation is modelled using one gas template at a time.
The likelihood of the results is compared and an ``index'' map constructed by selecting the maximum likelihood for each region.
The resulting composite gas map is used to analyse the simulation for the end result.

Tallying the per region likelihood comparisons, the correct gas map is selected most frequently for about half of the pixels.
For the pixels where the incorrect gas map is selected, it is in more than 80\% of the cases strongly correlated with the correct map in the regions, having Pearson's r-coefficient $>$0.9.
For those with smaller correlation, in about two thirds of the cases the incorrectly selected map is used in a neighbouring region in the correct map.
This still leaves $\sim$10 regions with an incorrect gas map that is not strongly correlated with the correct one, and not selected because of neighbouring regions.
For most of these, none of the maps is strongly preferred and the correct gas map is selected frequently, even though it is not the most frequent.
The reason for the method failing in those particular pixels is not clear, but it is likely caused by the correlations that are imposed in the analysis through the GPs.

To estimate the effect of these biases, Fig.~\ref{fig:best_fit_gas_simulation} shows the ratio of the analysis mean of the two emission components (structured, smooth) over the simulation truth.
If there was no bias, these maps should show no structure.
However, there are clear biases visible at the level of up to 20\%.
This should be compared to the standard deviation of the results, which is of the order of 10\% for both components.
This means that the bias is larger than the statistical power of the results, but only slightly.
The biasing is predominantly toward lower latitudes at the boundaries of our regions of interest, where the gas distribution is more uncertain and a larger fraction of the total intensity.
Bias in the recovered intensity of the smooth component is therefore dominated by low latitude emission.
The correlations for the lowest energy bin using Eq.~\ref{eq:gas_smooth_correlator} are shown in Fig.~\ref{fig:best_fit_gas_correlation}.
The correlations are overall small and only marginally larger than for simulations using a perfect gas map.
Compared to the upper panel of Fig.~\ref{fig:Bumps_correlation} the correlations are negative in this case and most of the largest deviations show signs of negative correlations.

We repeat the simulations with additional random permutations of the input truth map.
While the details change, the size and location of the bias is very similar between different simulation analyses.
We conclude that our method is able to reliably discriminate fluctuations that are larger in magnitude than few tens of percent, while smaller perturbations are susceptible to biases in the analysis.

\section{Results} 
\label{sec:results}

%Will be a figure set showing all models
\begin{figure*}
    \centering
    \includegraphics{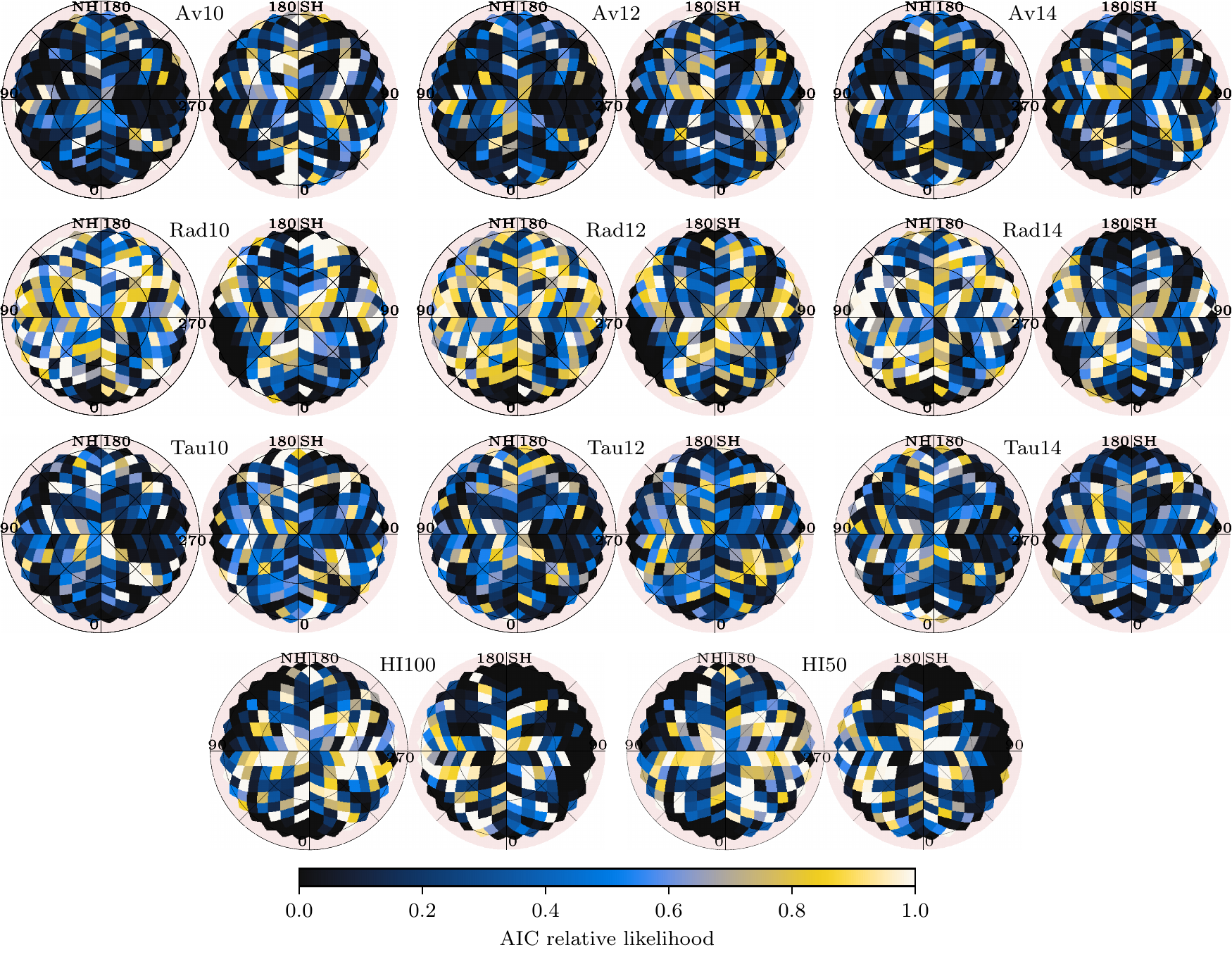}
    \caption{The AIC relative likelihood of the best-fit model for all of our input gas templates.
    The gas templates are labeled above each panel, with each gas tracer in a separate row and the conversion parameters in each column.
      The panels show the probability that the results of each simulation analysis give the best of the 11 models for each pixel.
    }
    \label{fig:likelihood_maps}
\end{figure*}

In the first phase of the work we analyse the LAT data (Sec.~\ref{subsec:data}) using each of the candidate gas column density maps (Sec.~\ref{subsec:model}) to establish the best-fit composite gas map.
For each run we require the $R$-statistic for the combined chains to be equal to 1 and visually inspect the posterior distributions and the trace of the chains.
These checks confirm that the chains are properly converged and 
that the posterior distributions for all parameters, except the hyper-parameters $\beta$ and $\rho$, are well approximated by normal distributions.
Therefore, we can use the posterior mean as an approximation for the maximum posterior, which gives a more robust result than using the maximum posterior value from the single best point within the chains.
The model, with the parameters determined this way, is then used to calculate the likelihood values for constructing the final best-fit gas map, as described in Sec.~\ref{subsec:adjust_gas}.
We only use the lowest energy bin to evaluate the likelihood, because it has the largest fraction of the intensity coming from the gas component and is therefore least sensitive to the details of the smooth component.
The maps of the AIC relative likelihood of the models are shown in Fig.~\ref{fig:likelihood_maps}.
White pixels represent the best model, while shades of yellow and light blue indicate models with high probability of being the best.
It is evident that while many regions show a clear preference for one map over others, there are many pixels without a clear distinction.
An example of the former case is toward the lower latitudes in the SH around $l\sim290^\circ$, while the polar regions are good examples of the latter case.  

\begin{figure}
    \centering
    \includegraphics{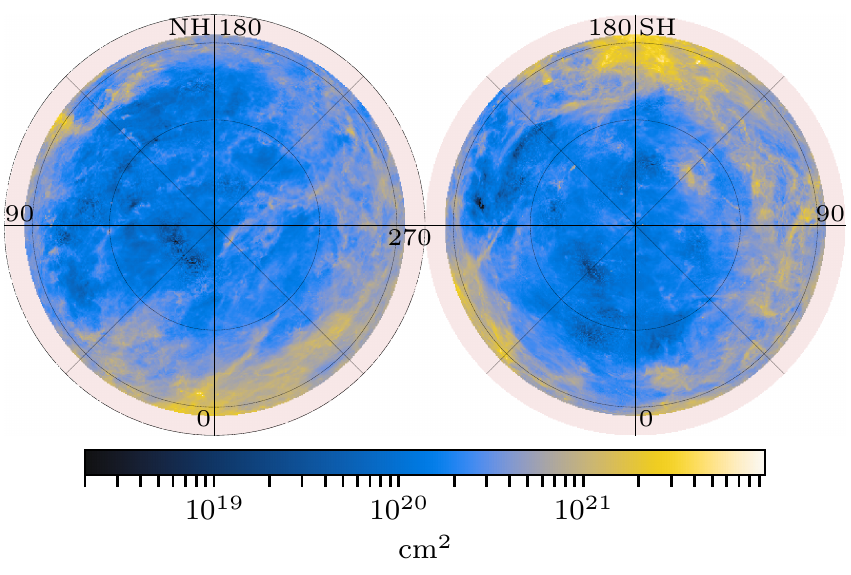}
    \caption{The composite column density map resulting from selecting the best-fit models shown in Fig.~\ref{fig:likelihood_maps}.}
    \label{fig:best_fit_gas_column}
\end{figure}

The overall best-fit gas map in the NH is the radiance map, and for the SH it is the $\tau_{353}$ map.
However, the best fit across different hemispheres is not always a strong indicator of the map being best for more localised regions.
In particular, the Rad12 map is much better on average than the HI50 map in both hemispheres, yet the HI50 map contains about twice as many pixels in the composite map used for the final analysis.
As a consequence, the exact size and location of the comparison regions may affect the final results to some degree.
The resulting gas map used for the final analysis, shown in Fig.~\ref{fig:best_fit_gas_column}, is therefore not unique.
Other gas maps with a similar level of optimisation and attendant per-pixel variation are possible, which depend on the details discussed above.
Such variations are expected to have an effect at a similar level as the simulations in Sec.~\ref{subsec:adjust_gas} (see Fig.~\ref{fig:best_fit_gas_simulation}) which are around 10\% to 20\%.

We caution also that the final column density gas map derived in this analysis should be interpreted with some care, because the absolute adjustment of the dust column density is uncertain and could very well vary between the different dust maps.
In case the absolute calibration of the dust maps is biased, the analysis will compensate for this by adjusting the emissivity.
Hence, in the final analysis any variations in the absolute column densities are corrected for by the emissivity term.
The effect of this is minimised by tuning the gas-to-dust ratio using the column density estimated from the optically thin \hi{} gas, but this only corrects the map globally over the whole sky.
If there is a significant variation in the gas-to-dust ratio over the sky, the determination of the emissivity will be affected, which will also in extreme cases affect the intensity of the smooth component.
As we show below, though, the lack of significant variations in emissivity and small anti-correlations between the emissivity and smooth components provides evidence that this is not a major issue, and the separation of components is sufficiently robust to support the results that we obtain.

The best-fit map shown in Fig.~\ref{fig:best_fit_gas_column} is extended outside the north/south hemisphere regions to account for the limited resolution of the LAT data.
We extrapolate by adding to the AIC maps rows of pixels outside the low-latitude boundaries.
  The values for the pixels are set to the mean of the adjacent pixels from inside the boundaries.

\begin{figure*}
    \centering
    \includegraphics{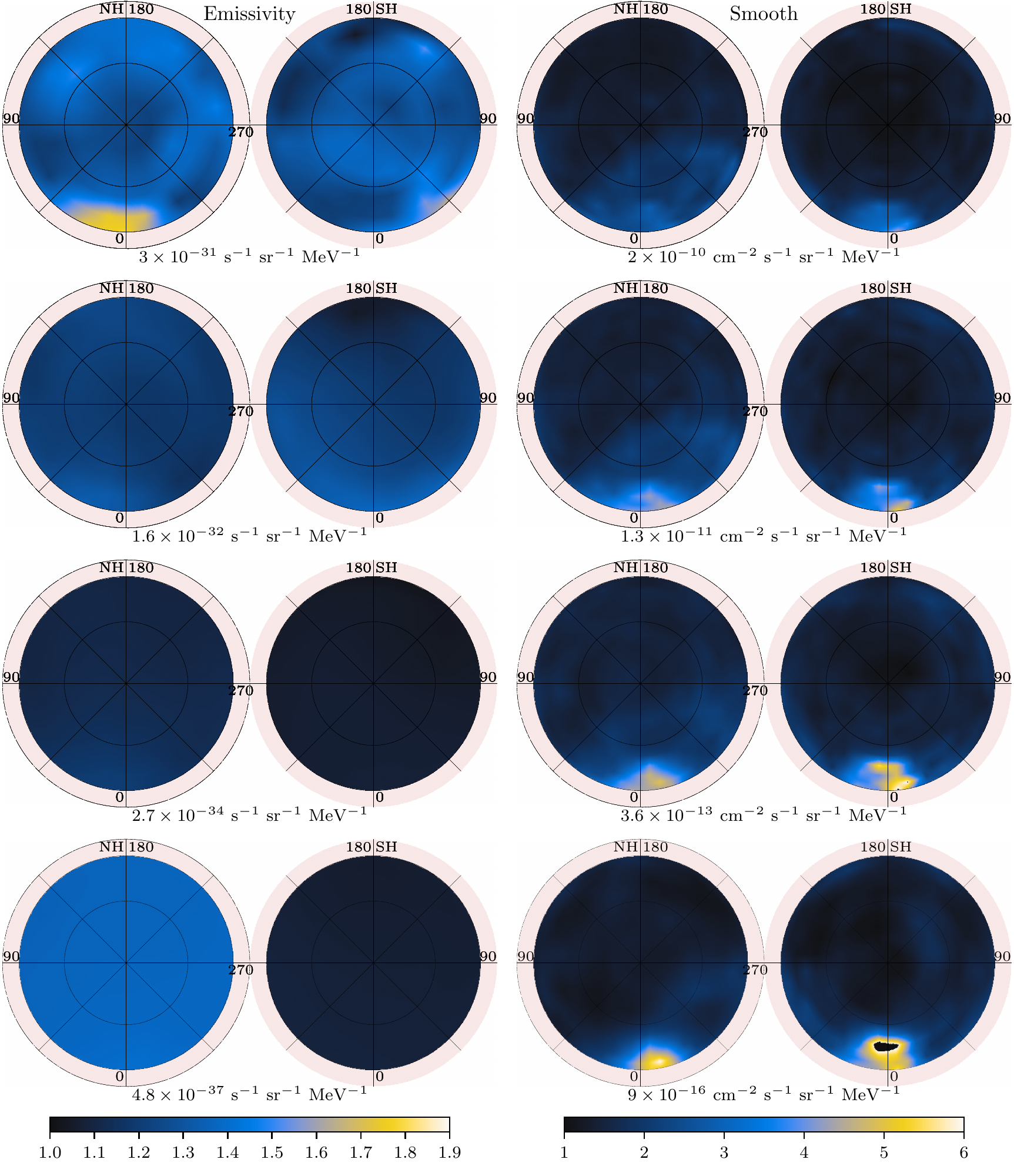}
    \caption{The directional gas emissivity (left column) and smooth component intensity (right column) resulting from analysis of LAT data using the gas map shown in Fig.~\ref{fig:best_fit_gas_column}.
      The maps are in orthographic projection with NH to the left and SH to the right.
      The directional emissivity and intensity are both evaluated at the geometric mean energy of each of the bins, from top to bottom: 1.7~GeV, 5.5~GeV, 24~GeV, and 240~GeV.
      The fractional colour scale is shown at the bottom of each column and the units of the maps are written below each panel.
    The black region about $l\sim0^\circ$ in the SH for the highest energy bin has pixels that saturate the colour scale.}
    \label{fig:Final_Results_all}
\end{figure*}

In the second phase we employ the best-fit column density map in the analysis of the LAT data to determine the distribution of the final directional emissivity and smooth intensity.
For the directional emissivity, its spectrum is initially set for each energy bin to the mean emissivity obtained over all the analyses performed in the first phase with the candidate column density maps. 
Similarly, the input spectrum for the smooth component is initially set to the mean spectrum of the smooth intensity obtained in the first phase.
Specifying the initial spectra of the individual components this way results in a better handling of the instrument response calculation when run through the GaRDiAn code, as described in Appendix~\ref{app:simulations}.
As before, convergence is checked by making sure the $R$-statistics as calculated by Stan are equal to 1 and manually inspecting the chains.

\begin{figure}
    \centering
    \includegraphics{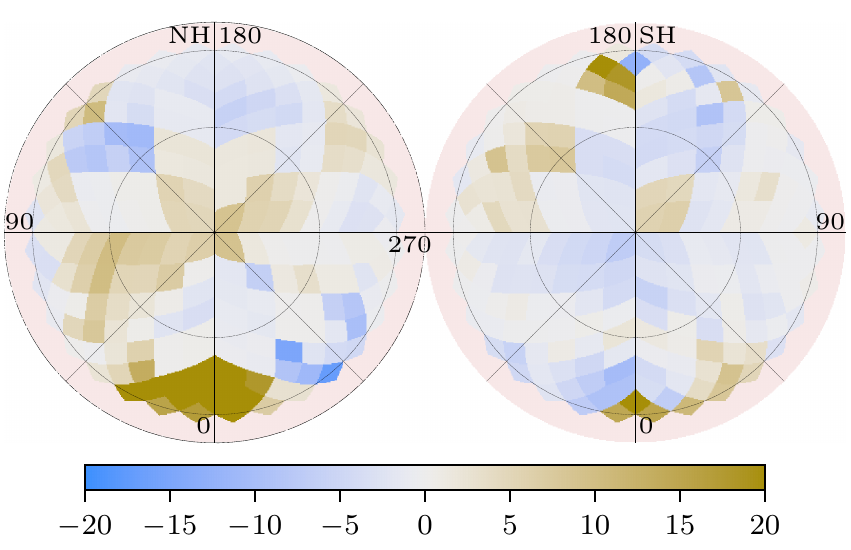}
    \caption{Correlation between the variation from the mean of the directional emissivity and smooth intensity according to Eq.~\ref{eq:gas_smooth_correlator} for the lowest energy bin using the results from the second phase of the LAT analysis.
      The maps are in Galactic coordinates and use the orthographic projection with the NH on the left and the SH on the right.
      The graticules are spaced $30^\circ$ apart in latitude and $45^\circ$ in longitude.
    }
    \label{fig:Final_Results_correlation}
\end{figure}

Figure~\ref{fig:Final_Results_all} shows the results of our analysis for the directional emissivity and smooth intensity\footnote{The results are also provided as fits files at \url{https://zenodo.org/record/4756707}.}.
The fluctuations for the directional emissivity are small, the range only being $\sim$2 times larger than the simulation results (Fig.~\ref{fig:best_fit_gas_simulation}).
Most should therefore not be considered significant.
The correlation map based on Eq.~\ref{eq:gas_smooth_correlator} for these results are shown in Fig.~\ref{fig:Final_Results_correlation}.
A few regions with negative correlations are apparent, hinting at systematic biases caused by the gas map despite our best efforts to minimise such cases.
For example, the peak around $l\sim140^\circ$ and $b\sim-40^\circ$ in the lowest energy bin corresponds to a dip in the smooth component.
It is likely an artifact of mismodelled gas structure and shows up as an anti-correlation.
The enhancement close to $l\sim45^\circ$ and $b\sim-30^\circ$ is, however, not associated with a strong dip in the smooth component, and shows up as a positive correlation.
It is therefore less likely to be an artifact.
But, the magnitude of this excess is still consistent with being a non-detection based on the analysis of the simulation fluctuations (Sec.~\ref{subsec:adjust_gas}).
The largest feature in the directional emissivity map is that visible between $\sim340^\circ$ and $\sim40^\circ$ longitudes in the NH.
This excess is apparent for the lowest energy bin, but disappears for the others.
Using \GP\ simulations (described in Appendix~\ref{app:simulations}) we estimate that an excess like this
occurs $\sim$1\% of the time, indicating that the statistical significance is $\lesssim$3$\sigma$.
The presence of this feature is not dependent on the exact gas map used for the analysis.
In other words, this excess exists at some level for all 11 gas column density maps used in this work, but it is not sufficiently significant to disambiguate from a fluctuation.

\begin{figure*}
    \centering
    \includegraphics{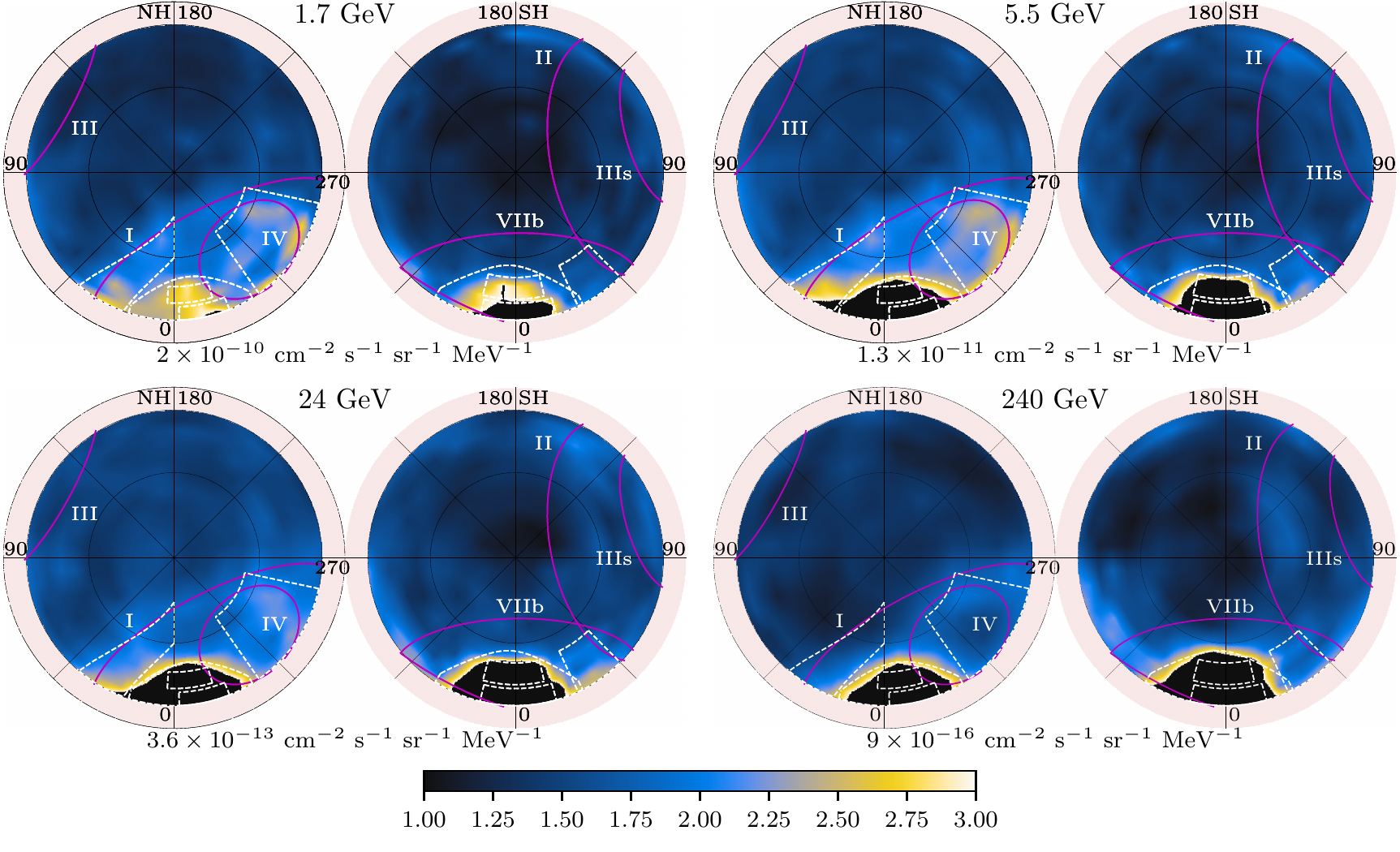}
    \caption{The intensity of the smooth component evaluated at the geometric centre of each of the energy bins used in the analysis.
      The range is reduced compared to Fig.~\ref{fig:Final_Results_all} and saturated pixels are shown as black.
      The magenta curves outline structures visible in radio synchrotron maps: the location and labels come from \citet{VidalEtAl:2015}.  The dashed white lines enclose regions that we determine the individual spectra (see Fig.~\ref{fig:results_spectrum_regions} below). The maps are shown in orthographic projection, with the NH to the left and SH to right, centred on the poles.  The fractional scale for the maps are identical, with the units given below each panel.}
    \label{fig:Loops_zoom}
\end{figure*}

For the smooth intensity component, there are extended emission features toward $l\sim0^\circ$ that appear at intermediate latitudes.
The feature in the NH is clearly offset to negative longitudes, while the one in the SH is offset to positive longitudes.
The shape of the feature in the NH appears to be fairly independent of energy, and is slightly dimmer than that in the SH.
The latter shows evidence of an energy dependent shape.
For the lower energy bins, the SH feature appears as two extended regions with the lower latitude one being brightest.
For the higher energy bins, the brightest part is centred more toward $l\sim0^\circ$ and higher latitudes.
Because the maps show the intensity interpolated from the GPs that are at a coarser angular scale than the data is given, careful interpretation is necessary.
The peaks are associated with the centres of the coarser pixels and consequently their specific maxima locations have some uncertainty.
There are, however, pixels in between the peaks with measurably lower intensities, so it is most likely that the shape for the SH regions is energy dependent.
The spectrum in the different regions is discussed with more detail below.

While the extended excesses toward $l\sim0^\circ$ are the most prominent ones, there are others that are less bright but still clearly visible.
In particular, there is strong evidence of emission from radio Loop I in the NH, extending in an arc from $l\sim40^\circ$ up to the pole and back down toward $l\sim300^\circ$.
To illustrate the lower surface brightness features, Fig.~\ref{fig:Loops_zoom} shows the smooth intensity component with a reduced intensity range.
Overlaid on the maps are traces of radio loops and spurs taken from \citet{VidalEtAl:2015}, selecting only those that are visible for the regions of sky considered by our analysis.
It is clear that there is evidence for emission from features labeled I and IV, while others are not as apparent in the map.
This is the first time that radio Loop~IV has been detected in high-energy \grays.
It is also clear that these features are softer than the overall background, because they are less prominent at higher energies.
The circular outline for radio Loop I seems to deviate from the \gray{} enhancement around $l\sim315^\circ$, where Loop IV becomes more prominent.
Loop IV also seems to be centrally filled, at least for the lowest two energy bins.

Other loops do not show emission over their entire range, but there seems to be an enhancement towards $l\sim35^\circ$ in the SH with $-50^\circ \lesssim b \lesssim -30^\circ$ at the base of Loops II and VIIb.
The spatial extent of this feature is smaller than that of Loops I and IV, but its magnitude is larger than expected from the systematic uncertainty estimated from the simulations.
Its spatial location also corresponds to a low-significance enhancement in the directional emissivity.
This feature is most prominent in the $5.5$~GeV and $24$~GeV energy bands with the emission at $1.7$~GeV and $240$~GeV considerably dimmer, indicating that its spectrum is harder than the background up to $\sim 50$-$100$~GeV at which point it softens.
Other features on the sky are much fainter and show up as light blue spots in the image.
These spots correspond to fluctuations at a level of 20\% to 30\%, which are similar to those observed for the simulations (Sec.~\ref{subsec:adjust_gas}).
They are susceptible to inaccuracies in the gas map that preclude attributing their origin for CR flux illumination by a localised source.
In particular, the faint emission at the base of Loop~II toward $l\sim160^\circ$ and $b\sim-35^\circ$ appears anti-correlated to some degree in Fig.~\ref{fig:Final_Results_correlation}.
Another feature is visible towards $l\sim -45^\circ$ in the SH close to the boundary of the region.
This feature is also anti-correlated with the directional emissivity, and, being constrained to the boundary, is susceptible to higher statistical and systematic uncertainty.
While there is some anti-correlation visible also for the Loop~IV region, the smooth component fluctuations in that region are so bright that gas-map inaccuracies are an unlikely explanation for its origin.

\begin{figure}
    \centering
    \includegraphics{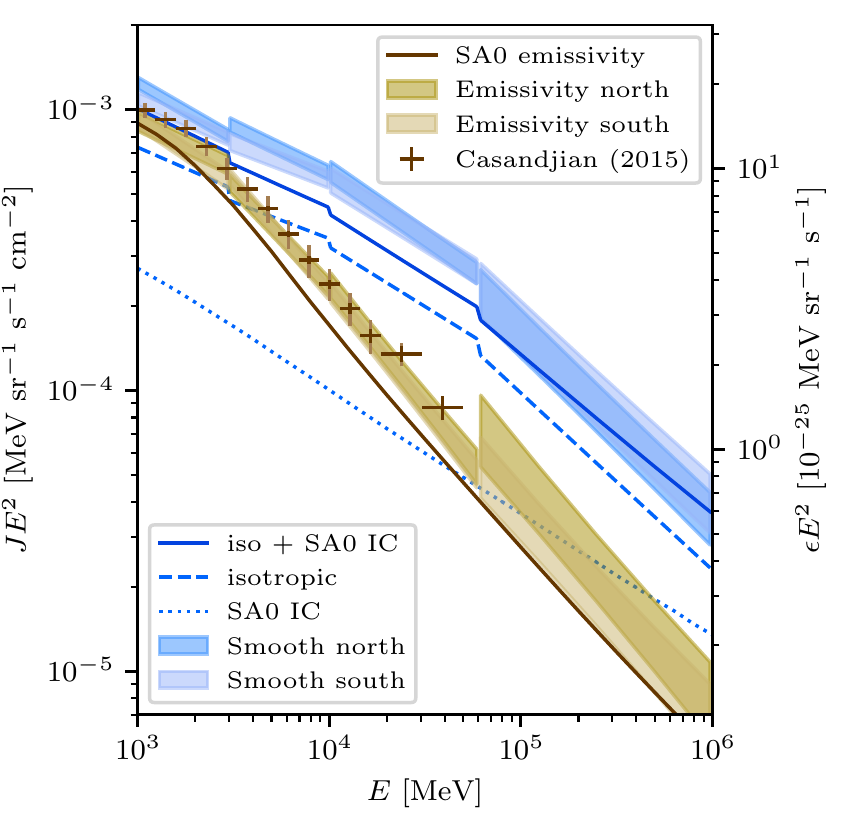}
    \caption{The average intensity of the smooth component (blue shaded regions) and the emissivity of the gas component (brown shaded regions) evaluated separately for the NH (darker colour) and SH (lighter colour), respectively.
      Also shown is the emissivity determined by \citet{Casandjian:2015}, the \GP\ predicted emissivity and IC intensity using the SA0 model from \citet{PorterEtAl:2017}, and the estimated isotropic spectrum in this analysis.
      The range $45^\circ < l < 270^\circ$ and $|b| > 30^\circ$ is used for the averaging.
      The left axis is for the intensities and the right axis for the emissivities.
      The axes are scaled so that the mean intensity of the gas component is approximately that shown in the left axis.       
    Differences in the mean gas column density in the NH/SH make accurate scaling impossible.}
    \label{fig:results_spectrum}
\end{figure}

The spectra for the spatially averaged smooth intensity and gas emissivity for $45^\circ < l < 270^\circ$ and $|b| > 30^\circ$ are shown in Fig.~\ref{fig:results_spectrum}.
This region was selected to exclude most of the features visible in the smooth component (Fig.~\ref{fig:Loops_zoom}).
Also shown is the emissivity spectrum derived from LAT data by \citet{Casandjian:2015}, and a prediction from \GP\ using the SA0 model from \citet{PorterEtAl:2017} that is tuned to reproduce the CR data.
The direction averaged gas emissivity obtained in this paper is consistent with the results of \citet{Casandjian:2015} above a few GeV.
It is somewhat higher than the gas emissivity predicted for the SA0 model, but the latter does not include the contribution from CRs heavier than Helium. 
This consistency with other estimates for the emissivity spectrum indicates that our separation of the \gray{} observations into smooth and structured contributions is successful.

The smooth component includes a contribution from residual CR background misclassified as \gray{s}, as well as \gray{} emission from unresolved point sources and the large-scale structure of the Universe.
This background emission is to first approximation isotropic.
To estimate its intensity we take for each energy bin the minimum value obtained in the smooth component for both hemispheres and show it as a blue dashed curve in Fig.~\ref{fig:results_spectrum}.
Also shown in the figure is the sum of our estimated isotropic component and the mean intensity of the IC component of the SA0 model as a solid blue curve.
It is clear that our direction averaged smooth intensity for this region is more intense than the combination of the SA0 IC and estimated isotropic emission.
However, direction averaged total \gray{} intensities over comparable sky regions for steady-state models normalised to the CR data that have different weighting for CR sources in the disc and spiral arms show $\sim$5--20\% variation, depending on energy \citep[e.g.,][]{PorterEtAl:2017,PorterEtAl:2019}.
The SA0 model higher-latitude intensities are generally lower than the other steady-state models.
A likely explanation that the smooth component is above the sum of the SA0 IC and derived isotropic intensities is that the large-scale properties for the steady-state CR source density are insufficiently optimised to reproduce the \gray{s} \citep[see also, e.g.,][]{2010ApJ...710..133A}.
Our determination of the isotropic component is also susceptible to statistical and systematic uncertainty.  
On one hand, the minimum value in the map will be biased low because of random fluctuations, underestimating the true isotropic intensity.
On the other hand, any contribution from IC emission from the old population of CR electrons is ignored, thus possibly overestimating the isotropic intensity.
The magnitude of the two effects are similar so they likely cancel each other.

\begin{figure*}
    \centering
    \includegraphics{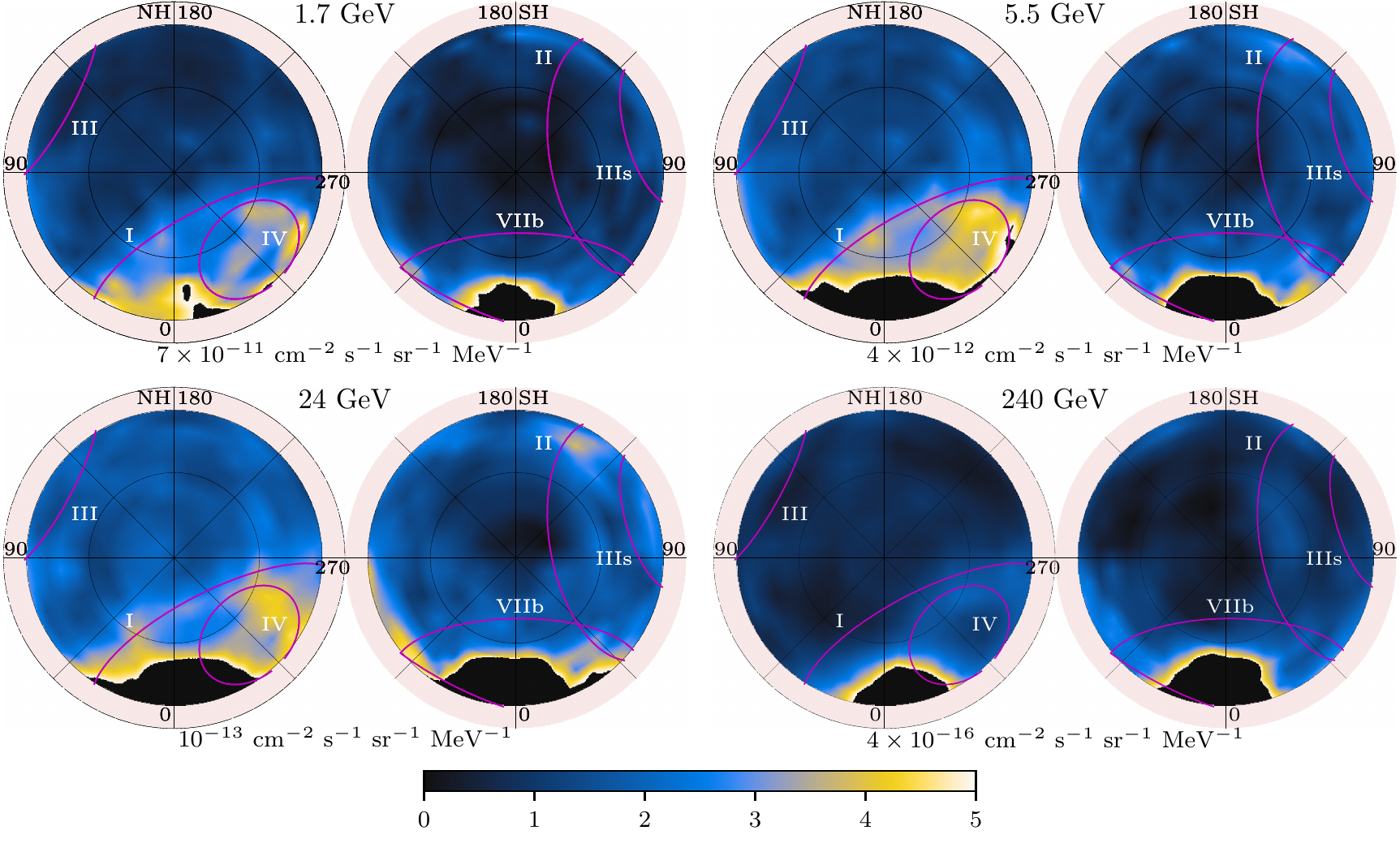}
    \caption{The intensity of the smooth component after subtracting an estimated isotropic background component, see text for details. For a description of the curves shown, see caption of Fig.~\ref{fig:Loops_zoom}. As for that figure, saturated pixels are shown as black.
    The fractional colour scale is adjusted so that a value of 1 corresponds to the mean of the IC emission in the SA0 model over the same region.}
    \label{fig:best_fit_smooth_subtracted}
\end{figure*}
  
The smooth component features shown in Fig.~\ref{fig:Loops_zoom} sit on top of the background, which is composed of the residual CRs, unresolved extragalactic sources, and the contribution from IC emission from the old population of CR electrons that have been accumulated over the lifetime of the Galaxy.
To highlight the features of the smooth component, we subtract from it our estimate of the isotropic component\footnote{The smallest value per energy bin of the smooth component.} and show the results in Fig.~\ref{fig:best_fit_smooth_subtracted}.
The IC emission from the old population of CR electrons cannot be determined from the observations, so any remaining contribution that is due to mismodelling of the IC is present in the isotropic-subtracted maps.
These are normalised such that a value of 1 corresponds to the mean intensity of the IC emission for the SA0 model predictions in each energy bin.
Compared to Fig.~\ref{fig:Loops_zoom}, the enhanced emission tracing the loop outlines appears to have a more consistent shape over the lower 3 energy bins, and the excess is visible even for the highest one, albeit with a low statistical significance.  

\begin{figure}
    \centering
    \includegraphics{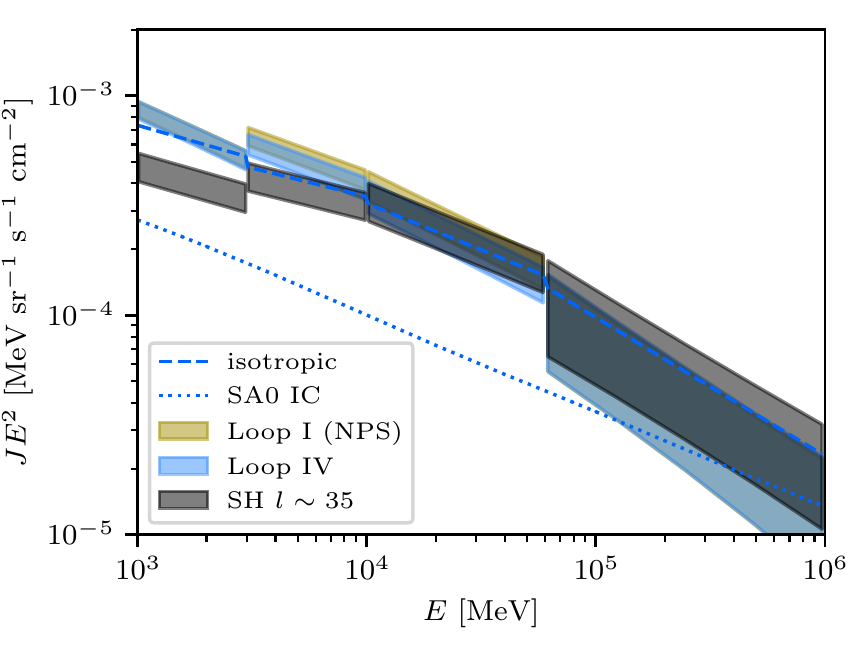}
    \includegraphics{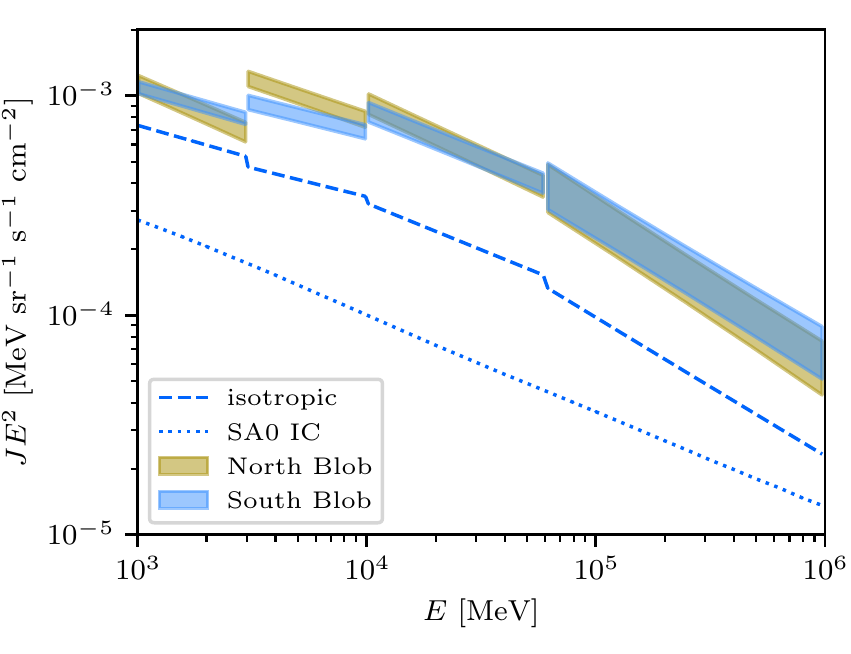}
    \includegraphics{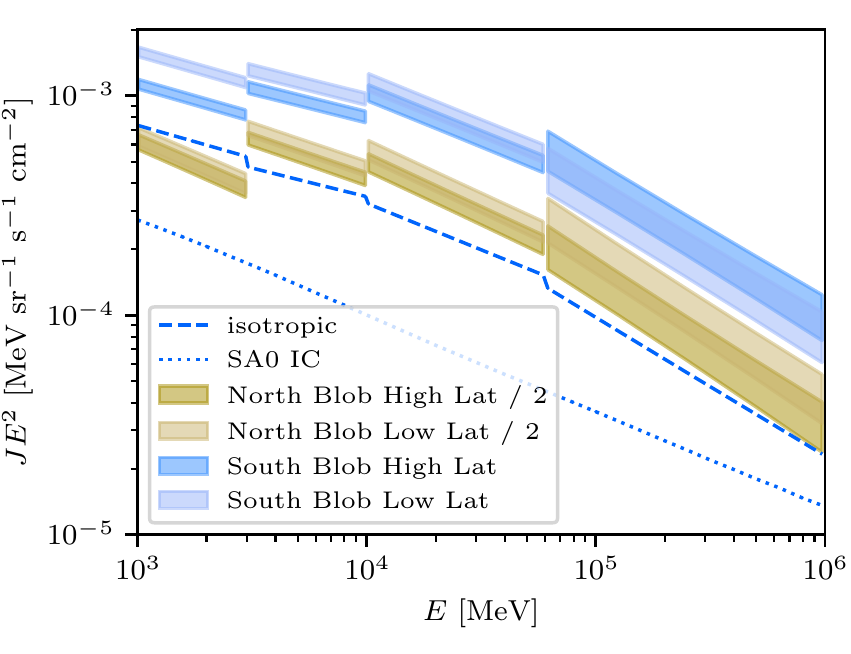}
    \caption{Spectral intensities following isotropic subtraction (see text for details) derived for the bright smooth component regions delineated by the white lines in Fig.~\ref{fig:Loops_zoom}.
      Panels: top, loop-associated; middle, NH/SH blob total; bottom, NH/SH blobs split into low/high-latitude sub-regions.
      Note that the NH blob sub-regions are scaled down by a factor 0.5 for clarity.
    All panels also show the estimated isotropic spectrum and the IC prediction from the SA0 model.}
    \label{fig:results_spectrum_regions}
\end{figure}

To expand on the characteristics for the bright features in the smooth component described above, we determine the spectral content for the individual features using the white dashed outline regions shown in Fig.~\ref{fig:Loops_zoom}.
The resulting isotropic subtracted spectral intensities are shown in Fig.~\ref{fig:results_spectrum_regions}.
For the features associated with the Loop I and IV overlays (top panel), the spectra are essentially the same and very similar to that of the isotropic component, while the feature visible at the base of Loops II and VIIb is clearly harder.
Meanwhile, the aggregate spectra for the NH/SH blobs toward $l\sim0^\circ$ are similar (middle panel), and generally harder than the loop-associated features, even the harder feature at the base of Loops II and VIIb.
However, when we separate into approximately comparable ``lower/higher'' latitude sub-regions (see Fig.~\ref{fig:Loops_zoom}), in the bottom panel we can see that the spectral characteristics for these blobs are different.
Specifically, the NH blob lower/higher latitude spectra are very close in shape.
On the other hand, the SH blob lower/higher latitude spectra are different.
The low-latitude region has a similar spectrum to that for the NH blobs, whereas the high-latitude region is noticeably harder than all other regions.

Note that the spectral shape within each of the coarser energy bins displayed in Fig.~\ref{fig:results_spectrum_regions} is determined from a global fit over individual bins in each hemisphere.
As such it cannot accurately reproduce emission that is softer/harder than the average within each energy bin.
This is evident for the spectra of the two blobs: they are significantly harder than the background, resulting in the edges of the coarse energy bins being discontinuous.

\section{Discussion}
\label{sec:discussion}

\begin{figure*}
    \centering
    \includegraphics{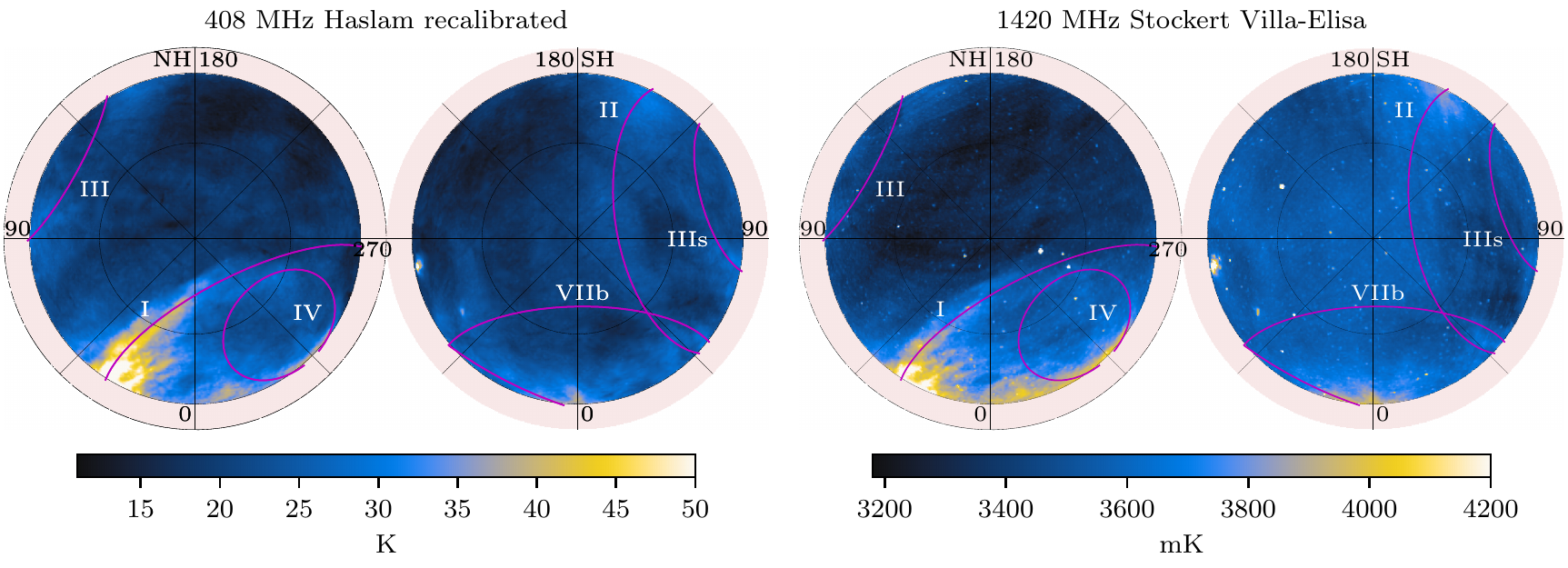}
    \caption{Example emission maps at radio frequencies. Left: 408~MHz map of \citet{1982A&AS...47....1H} as reprocessed by \citet{2015MNRAS.451.4311R}. Right:  1.4~GHz map of \citet{1986A&AS...63..205R,2001A&A...376..861R}.  The colour scale is adjusted such that the emission from the NPS has similar brightness in both maps.}
    \label{fig:radio_maps}
\end{figure*}

Comparing the spatial distribution of the smooth intensity component, which is most likely due to IC emission from CR electrons, with radio maps at MHz/GHz frequencies (shown in Fig.~\ref{fig:radio_maps}) reveals both similarities and differences.
The North Polar Spur (NPS), the low latitude part of Loop~I, is clearly bright in both energy ranges, at a similar level above the background, while the fainter loops designated III and IIIs are not visible in \gray{s}.
For the \gray{s}, there are hints of emission from Loop~II and~VIIb in the SH toward $l\sim35^\circ$ that can be seen for both the 5.5~GeV and 24~GeV bins, but the significance is low.
The radio maps show no significant feature in that region.
Meanwhile, Loops~I and~IV are very bright in \gray{s} and clearly visible up to the highest energies considered in this paper.
The \gray{} intensity of Loop~IV is comparable to that of Loop~I and the NPS, and the intensity of Loop~I is nearly constant over the entire structure.
Contrasting, the NPS is the brightest feature in the radio maps with the other loop features generally having similar intensities.

It is generally thought that these loops and spurs are nearby and old SNRs, where the shock compressed magnetic field is able to contain and accelerate an electron population that is more dense than that of its surroundings.
According to Eq.~(3.13) from \cite{MertschSarkar:2013}, the spectrum of the electrons in a shock with a compression factor of $\eta$ is given by
\begin{equation}
    n'(p) \approx \frac{2}{\sqrt{\eta}} n\left(p\sqrt{\frac{3}{2\eta}}\right),
\end{equation}
where $n(p)$ is the spectrum in the ISM.
Assuming a power-law spectrum $n(p)\propto p^{-\gamma}$, the ratio between the two can be estimated as
\begin{equation}
  \frac{n'(p)}{n(p)} \approx \frac{2}{\sqrt{\eta}}\left(\frac{3}{2\eta}\right)^{-\gamma/2},
  \label{Eq:ratio}
\end{equation}
which is independent of momentum and its value is around 6 for the canonical values of $\eta=6$ and $\gamma=3$.
Assuming that the ISRF density does not vary much across the shock, the intensity of the IC emission should also be increased by the same factor in the shock.
For the lower energy radio/synchrotron radiation, the emissions are also modified by the enhancement of the magnetic field in the shock by a factor of $\eta$, leading to an even larger increase in the emission than estimated from the increased electron density.

However, the comparison between the radio/synchrotron and the GeV IC emissions is not direct, because of the differing CR electron energies producing the emissions for the respective ranges.
The energy of electrons producing the 1~GHz synchrotron emissions is $\sim$10~GeV for a magnetic field of a few $\mu$G.
Meanwhile, the $\sim$10~GeV IC \gray{s} are produced by $\gtrsim$70~GeV electrons upscattering the starlight and infrared components of the ISRF, which are the major target photon fields over the energy range considered in this paper.
The lack of emission from some of the loops therefore indicates that they cannot efficiently confine the higher-energy electrons, or that they are produced only by an increased magnetic field strength, which is not reflected in an increased CR electron density.
The results presented here therefore imply that Loops~I and~IV are efficient in confining or accelerating CR electrons, with Loop~IV capable of confining/accelerating electrons up to at least TeV energies to explain the enhancement in the highest-energy bin.
For Loop~IV this is surprising, because it does not stand out for the radio/synchrotron emissions.

The shock compressed magnetic field is also expected to increase the energy of the CR protons and nuclei.
The lack of a spatially coincident increase in the directional emissivity of gas indicates that structure for these shocks is not well captured by the column density map used for our analysis (Fig.~\ref{fig:best_fit_gas_column}).
If the gas in the shocks is mostly ionised then they would not be evident in the emissivity because we have utilised neutral gas maps to trace likely hadronic emissions.
Tracers of ionised gas, such as observations of H$_\alpha$ emission \citep{Finkbeiner:2003}, however, do not show any structure that is spatially associated with the radio loops.
The intensity from the loops shown in Fig.~\ref{fig:results_spectrum_regions} is approximately 3 times that of the predicted SA0 IC intensity giving a compression factor in Eq.~\ref{Eq:ratio} of $\eta\sim3$.
This can be employed to place a limit on the column density of gas in the loops.
Applying the same compression factor to the emissivity spectrum from Fig.~\ref{fig:results_spectrum} at 1~GeV results in $\epsilon E^2 \sim 3 \times 10^{-24}$~MeV~sr$^{-1}$~s$^{-1}$.
The intensity of the loops at the same energy is $JE^2 \sim 7\times 10^{-4}$~MeV~sr$^{-1}$~s$^{-1}$~cm$^{-2}$.
Requiring that the intensity of any gas component is $\lesssim$30\% of the intensity of the loops in the lowest energy bin, results in a column density of gas $\lesssim$$10^{20}$~cm$^{-2}$.
If it were larger there would be signatures of the hadronic emission in the intensity spectra for the loops.

Note, that the above discussion refers to any gas that {\it is not} following the neutral tracers used in this analysis.
If instead the gas in the loops is contained in the neutral structured maps (Fig.~\ref{fig:best_fit_gas_column}), the CR proton enhancement within the structures is actually smaller than that for the electrons, otherwise there would be a signal in the directional emissivity.
Or, alternatively, only a fraction of the neutral gas column density is coincident spatially with the location of the CR injection.
Then, the structured component signal could be low also because of the sparsity of neutral target density for producing the \gray{s}.
If that is the case, the origin of the emission is likely well above the gas layer of the Galaxy, placing them at $\gtrsim100$~pc height above the Galactic plane.

Recent results from the eROSITA X-ray telescope confirm the presence of significant X-ray emission in the direction of radio Loop~I and a counterpart in the SH at negative longitudes~\citep{2020Natur.588..227P}.
Comparison of the X-ray emission (their Fig.~2) with the intensity of the smooth component derived in this paper again shows some similarities and differences, as with the radio/low-energy synchrotron.
The Loop~I structure extends to higher latitudes and all the way to negative longitudes in X-rays, similar to that in \grays.
The Loop~IV structure visible in \grays\ is, however, absent in the X-rays.
There is a strong feature in the X-rays visible toward $l \sim -40^\circ$ that is extended in latitude down to $b\sim -45^\circ$, while the feature observed in \grays\ at the similar location is restricted to $b>-35^\circ$ and extended in longitude.
The \gray\ feature is susceptible to inaccuracies in the gas map toward those directions, though.
There is no clear feature observed in the X-ray emission in the SH at $l\sim35^\circ$ where there is evidence of emission in \grays.

According to \citet{2020Natur.588..227P}, the X-ray emission is interpreted to come from giant X-ray emitting spherical shells of hot and tenuous plasma.
Co-location of the \gray\ emission with the X-ray emission in this model would require particle acceleration within the plasma, either electrons for IC emission, or hadrons.
Similar interpretations for the FBs have been invoked \citep[e.g.,][and references therein]{2018Galax...6...29Y}.
For the FBs, the former explanation is difficult because of fast cooling of the electrons, while the latter requires a substantial energy budget.
  Because of the larger shape of the X-ray bubbles and their larger distance from the Galactic plane, the difficulties are even more severe.
%The \gray\ observations therefore point to a local origin for the Loop~I emissions.
The spectra of the loops, shown in Fig.~\ref{fig:results_spectrum}, are softer than that of the FBs, but there is no clear evidence of a spectral cut-off as would be expected from increased cooling.
  However, due to the the range of the last energy bin in our analysis (60--1000~GeV), we cannot rule out a spectral cutoff at energies $\gtrsim$100~GeV.

One of the arguments of \citet{2020Natur.588..227P} in favour of the giant X-ray shells being associated with the FBs is the apparent symmetry of the results, both in the NH/SH and along the $l=0^\circ$ meridian.
Our results do not provide evidence in favour of this symmetry, and in fact are somewhat in tension with this interpretation because the \gray{} features that are recovered by our analysis are not symmetric about the Galactic plane nor $l=0^\circ$ meridian.
The correspondence between the X-rays and \grays\ in the NH argues for a common origin of the emissions in this region of the sky, while a physical connection with the SH is less certain.
Visual comparison of Fig.~2 from \citet{2020Natur.588..227P} indicates our method should be sensitive to emission in the SH tracing the X-ray emission.
To make the SH results compatible with those in the NH requires a ratio increase
%However, a possible increase in the ratio
between the X-ray and \gray{} emission by a factor of $\gtrsim$2, but the physical explanation for such a factor for this region if there is a common origin is not clear.
%in the SH would make the results compatible.
It also remains a possibility that the \gray\ and X-ray emissions from the regions around Loop~I are not at all correlated, although it seems unlikely that two large structures overlapping on the sky do not have a physical connection.
%While the X-rays show a faint southern component with a clear lower-latitude feature (see Fig.~2 from  \citet{2020Natur.588..227P}), there is nothing evident at \gray{} energies.
So while we cannot rule out that there is a common origin, we believe our results currently argue against strong claims that this is the case for the emissions in the two hemispheres, and that they are necessarily associated with past activity toward the Galactic centre (GC).

%, and the lack thereof in the SH, indicates that the features observed in X-rays in each hemisphere {\bf may not be} related.
%{\bf A possibility is that the \gray\ and X-ray emissions from the regions around Loop~I may not be correlated, but arguably} it is unlikely that two large structures overlapping on the sky do not have a physical connection.

The spectral shape that we obtain for these features generally agrees much better with a leptonic, rather than hadronic, origin, at least if they are local.
If they are not local, then the spectral shape cannot say much about the origin of the emissions because the underlying particle distribution is not well constrained by other observations.
Arguments can, however, still be made in favour of a leptonic origin of the emission if it originates from large distances. 
For the spectral intensities that we obtain, either a high target density or high CR flux is necessary in the regions covering the features.
In the hadronic scenario, the former is ruled out by lack of features in the neutral gas traced structure component, and the very low density estimated from the X-ray data \citep[$\sim$$10^{-3}$~cm$^{-3}$;][]{2020Natur.588..227P}.
Assuming this density and an integration path length $\sim$10~kpc, the column density is $\sim$$3\times10^{19}$~cm$^{-2}$.
Consequently, the mean emissivity, and hence the mean CR density, needs to be more than 15 times that of the local ISM for the lowest energy bin in our analysis, where we expect most of the CR energy density to reside.
Given that the region has a size of 10 kpc, which is a volume of about a quarter of the galaxy, the energy in CRs is at least a few times larger than in the entire Galaxy.
It seems difficult to create sufficient CRs and then confine them in the sky regions covered by the loops if their origin is far away along the line-of-sight.

The distances to the radio loop structure, in particular the NPS and radio Loop~I have been the subject of debate in the literature since their discovery \citep[e.g.,][]{2018Galax...6...56D,2018Galax...6...27K}.
The large size of the features, and their asymmetry around the GC, indicates a local origin, possibly associated with a massive star-formation event in the Sco-Cen OB association located at a distance of $\sim$120~pc \citep{VidalEtAl:2015}.
However, absorption measurements of the X-ray spectra and the temperature of the X-ray emission argue for a distance more than a few hundred pc.
The lack of any signature in the directional emissivity towards the direction of Loops I and IV in our results indicates that the emission is taking place well above the Galactic plane, above most of the gas layer.
Assuming a physical radius of 200~pc, \citet{MertschSarkar:2013} have computed distances to Loops~I-IV (see their Table 2), with Loops~II and~IV having distance and centres $|b| > 30^\circ$ indicating their origin is likely $\gtrsim$100~pc from the Galactic plane.
For example, the Loop~IV centre is estimated $\sim$590~pc away, placing it at a height $\sim$440~pc.
Because the ISRF density falls off out of the Galactic plane much slower than that of the gas \citep[][]{PorterEtAl:2017,2018ApJ...856...45J}, the targets for IC emission are much more abundant at such heights.
This accords with the predictions of \citet{PorterEtAl:2019} who showed that such high-latitude discrete sources should be brightest in IC emissions.

Of the features observed in the sky, the one in the SH at $l\sim35^\circ$ is most similar to those predicted by \citet{PorterEtAl:2019}.
It has a spectrum that is harder than the surrounding region, and it potentially shows evidence of increased emission in both the gas and smooth components, although the former is not very significant.
The lack of corresponding feature in the radio and X-rays, as well as the spectral turnover observed in the highest energy bin, indicates that the emission might be caused by particles diffusing away from a not so recent event that happened well above the Galactic plane.
Examining the ATNF Pulsar Catalogue (version 1.64)\footnote{https://www.atnf.csiro.au/research/pulsar/psrcat/} \citep{2005AJ....129.1993M}, the pulsar B2045-16 \citep{1968Natur.219..689T} is located at a distance of $\sim$1~kpc in that region with a characteristic age of $\sim$$3\times10^6$~yrs.
This corresponds to a cooling timescale for the electrons with energies $\sim$300~GeV assuming an energy density of 1~eV~cm$^{-3}$ for the photon and magnetic field, in agreement with the observed spectral properties.

The bright NH/SH blobs toward $l\sim0^\circ$ are approximately coincident with high-latitude regions covered by the FBs.
The spectral evolution across the features that we obtain is consistent with prior analyses, where the southern spectra harden with increasing latitude \citep[e.g.,][]{2014A&A...567A..19Y}.
This is contrary to initial findings that suggest invariant spectral characteristics for the FBs \citep[e.g.,][]{2010ApJ...724.1044S,2013PDU.....2..118H}.
The spectral intensities that we obtain (Fig.~\ref{fig:results_spectrum_regions}) are generally higher for both NH and SH regions than obtained by prior works.
However, the earlier analyses employed \GP-generated gas-related ($\pi^0$-decay and bremsstrahlung) and IC templates for a variety of 2D CR propagation models.
Their results are therefore reporting the bubble spectral characteristics that are residuals after subtracting the emissions from these Galaxy-wide emissions.
For our work, we have employed an optimised structured (gas) component and imposed only a spatial connectedness requirement (via the GPs) for the smooth component.
The spectral intensity that we obtain, after determining the minimal isotropic contribution, likely still contains a pedestal of IC emission due to the `sea' CR electrons that is combined with localised enhancements in these regions.
However, its determination is model-dependent, as described earlier.

Even with this pedestal, the spatial distributions for the NH/SH blobs are clearly asymmetric and localised. 
The spatial distributions $|b| > 30^\circ$ that we recover differ in detail from other analyses \citep[e.g.,][]{2014ApJ...793...64A,2014A&A...567A..19Y,2015A&A...581A.126S}.
But, the general shapes, including asymmetries, are similar with those obtained by the prior works: the northern feature has orientation tilted to negative longitudes while the southern one is to positive longitudes \citep[see, e.g., Fig.~23 of][]{2014ApJ...793...64A}.
The overlap with the Loops I and IV for the NH blobs precludes clean separation, and it is difficult to identify boundaries between the individual features.
The emissions from these two features have often been neglected, or approximated using templates based on simple geometrical models \citep[e.g.,][]{2010ApJ...724.1044S} or the 408~MHz radio map \citep[e.g.,][]{2014ApJ...793...64A}.
In particular, the emission from Loop~IV as a distinct feature has not been included explicitly in prior analyses, which could affect the determination of the morphology of the FB in the NH for energies $\lesssim$100~GeV.
For the SH, the emission surrounding the saturated part of the intensity scale (Fig.~\ref{fig:best_fit_smooth_subtracted}) has variable spatial morphology that is energy dependent.
From the edge of the saturated region to where the emission is comparable to the SA0 intensity, the falloff on average is approximately linear over $\sim$$10^\circ$.
This range corresponds to the spatial resolution of the underlying GP, which also has a correlation length of that order.
We are therefore not able to make a definite conclusion whether the NH/SH blobs have abrupt boundaries, as claimed by prior analyses of the FBs \citep[e.g.,][]{2010ApJ...724.1044S}.

\section{Summary}
\label{sec:summary}

In this work we used GPs to model the spatial distribution of the directional gas emissivity and the intensity of a smooth component of the high-latitude ($|b|>30^\circ$) \gray{} sky between 1 GeV and 1 TeV.
The GPs impose a minimal spatial connectedness requirement, but otherwise allow for substantial flexibility to trace spatial structures in the \gray{} data.
  We thoroughly tested the methodology with simulations.
  Systematic uncertainties caused by inaccuracies in the distribution of interstellar gas, which can be problematic when analysing large regions of the \gray{} sky, are controlled by developing a dedicated column density map based on linear interpolation of various gas tracers.
Using this optimised gas map, the fractional systematic uncertainty due to the imperfectly known gas column density distribution on the sky is $\lesssim$$\pm20$\%.
The results that we obtain for the directional emissivity are consistent with being constant within our estimated statistical and systematic uncertainty, apart from a region toward $l\sim 20^\circ$ and $b \sim 35^\circ$ (the approximate centre of the broad region shown in upper left panel of Fig.~\ref{fig:Final_Results_all}), where an excess is found with a $\lesssim$3$\sigma$ significance.

The brightest structures that we detected with the smooth intensity component are hard emission features located towards the GC that are offset to negative longitudes in the NH, and positive longitudes in the SH.
These features are spatially consistent with the higher latitude emissions attributed to the FBs, and our results are qualitatively consistent with other work.
The emissions that we obtain are spectrally harder than the surrounding background, with the average spectral features in the NH and SH similar.
The feature in the SH shows evidence of spectral evolution, with the emission at $b\sim35^\circ$ being softer than the emission at $b\sim45^\circ$.
No such evolution is evident for the feature in the NH.

We also detect significant emission from the radio Loops~I and~IV.
The intensities of these individual loops in \gray{s} are comparable, and they extend to $\gtrsim$100~GeV energies.
Such detection contrasts with the situation in X-rays and radio, where Loop~IV is much dimmer than Loop~I.
The hard spectrum of the \gray{} emission suggests a leptonic origin, and indicates these structures are located at a distance of $\gtrsim$500~pc, given their non-detection in the directional emissivity.
We also detect a marginally significant signal from the bases of Loops~II and VIIb towards $l~\sim 35^\circ$.
This enhancement is of similar magnitude as that of Loops~I and IV, but its spectrum is significantly harder.
There is also a faint signal detected toward this direction in the directional emissivity, but it is consistent with expected fluctuations from inaccuracies in the gas map.
Nevertheless, this emission could come from a recent and local CR injection event.

Our results support our earlier predictions that the high-latitude \gray{} sky encodes recent CR source injection and propagation history.
While our analysis is not able to provide definitive spatial localisation for the features that we detect, there is a good chance that they are relatively nearby (i.e., not emanating from the GC) given the asymmetric spatial distributions and spectral characteristics.
If so, there is a possibility that they may produce spectral features in the local CR data.
Modelling of such is beyond the scope of the present investigation, but our work provides evidence for the possible connection between the high-latitude \gray{s} and the high-energy CR data.

\acknowledgments

The \textit{Fermi} LAT Collaboration acknowledges generous ongoing support from a number of agencies and institutes that have supported both the development and the operation of the LAT as well as scientific data analysis.
These include the National Aeronautics and Space Administration and the Department of Energy in the United States, the Commissariat \`a l'Energie Atomique and the Centre National de la Recherche Scientifique / Institut National de Physique Nucl\'eaire et de Physique des Particules in France, the Agenzia Spaziale Italiana and the Istituto Nazionale di Fisica Nucleare in Italy, the Ministry of Education, Culture, Sports, Science and Technology (MEXT), High Energy Accelerator Research Organization (KEK) and Japan Aerospace Exploration Agency (JAXA) in Japan, and the K.~A.~Wallenberg Foundation, the Swedish Research Council and the Swedish National Space Board in Sweden.

Additional support for science analysis during the operations phase is gratefully acknowledged from the Istituto Nazionale di Astrofisica in Italy and the Centre National d'\'Etudes Spatiales in France. This work performed in part under DOE Contract DE-AC02-76SF00515.

\GP\ development is partially funded via NASA grant NNX17AB48G.
Some of the results in this paper have been derived using the 
HEALPix~\citep{2005ApJ...622..759G} package.

\software{Stan (https://mc-stan.org), 
  Astropy \citep{astropy:2013, astropy:2018},
  HEALPix \citep{2005ApJ...622..759G},
  NumPy \citep{harris2020array}
}

\bibliography{hlpaper}{}
\bibliographystyle{aasjournal}

\appendix

\section{Simulations}
\label{app:simulations}

\begin{figure}
    \centering
    \includegraphics{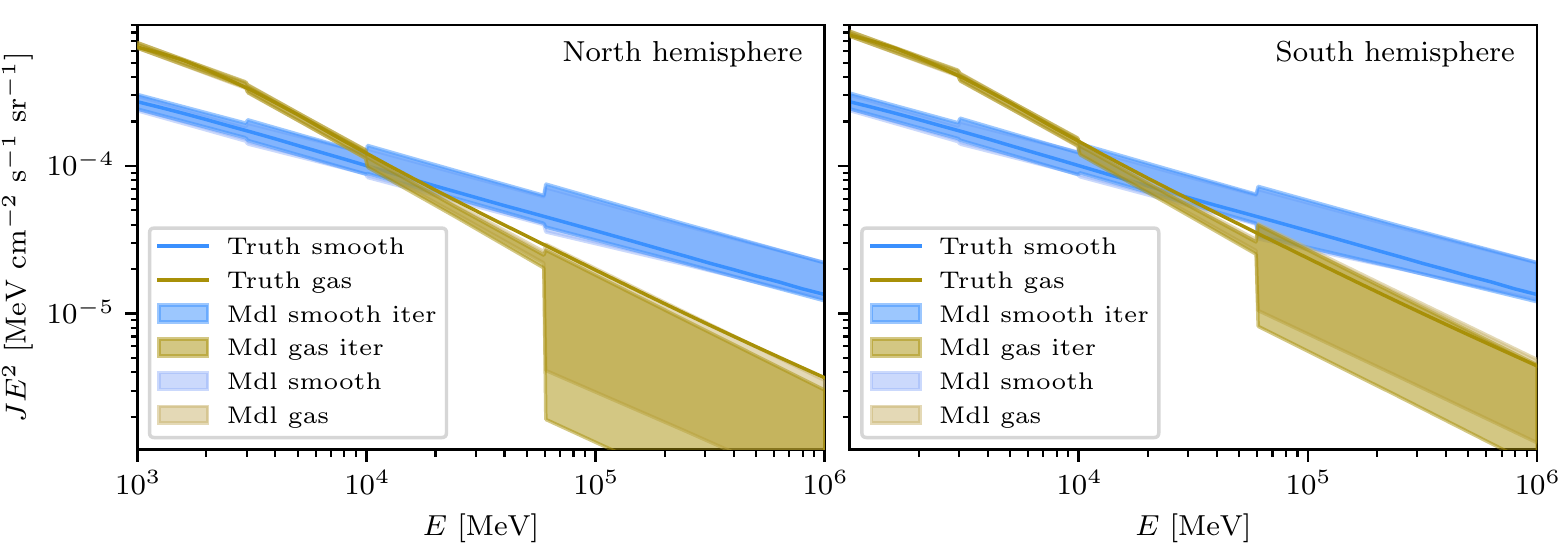}
    \caption{The intensity spectrum of the gas and smooth component and 68\% confidence regions after analysing the 2000 simulations that are based on a \GP\ run.
      The brownish and blueish shaded areas show the spectrum of the gas and smooth components, respectively.
      The lighter shades represent the results without any iterations, while the darker shades represent the results after a single iteration of the analysis.  The solid curves show the model input spectra.
      Left/right panel: NH/SH model and analysis results, respectively.}
    \label{fig:GALPROP_spectrum}
\end{figure}

The first simulation is based on a \GP\ calculation, the steady-state, azimuthally-symmetric CR source model designated SA0 and described by \citet{PorterEtAl:2017}.
This model provides a realistic spectral shape for the gas emissivity and smooth intensity because it is optimised to reproduce the CR data.
This simulation and analysis tests if the assumption of a power-law spectrum within each of the four spectral bins is good enough to capture the expected spectral shape, and if our assumptions of fitting in counts space have a strong effect on our results.
Figure~\ref{fig:GALPROP_spectrum} shows the resulting spectrum of the two components from our analysis, averaged over the 2000 simulations.
Also shown is the standard deviation of the spectrum, which indicates the possible statistical accuracy of the spectrum.
To test the effect of fitting in counts space, a single iteration of the analysis is performed by adjusting the spectral shape of the templates used to match the average of the simulations.
The figure illustrates that the changes due to the more accurate convolution with the IRFs per-likelihood evaluation are much smaller than the statistical uncertainties, and such iterations are not needed.

\begin{figure}
    \centering
    \includegraphics{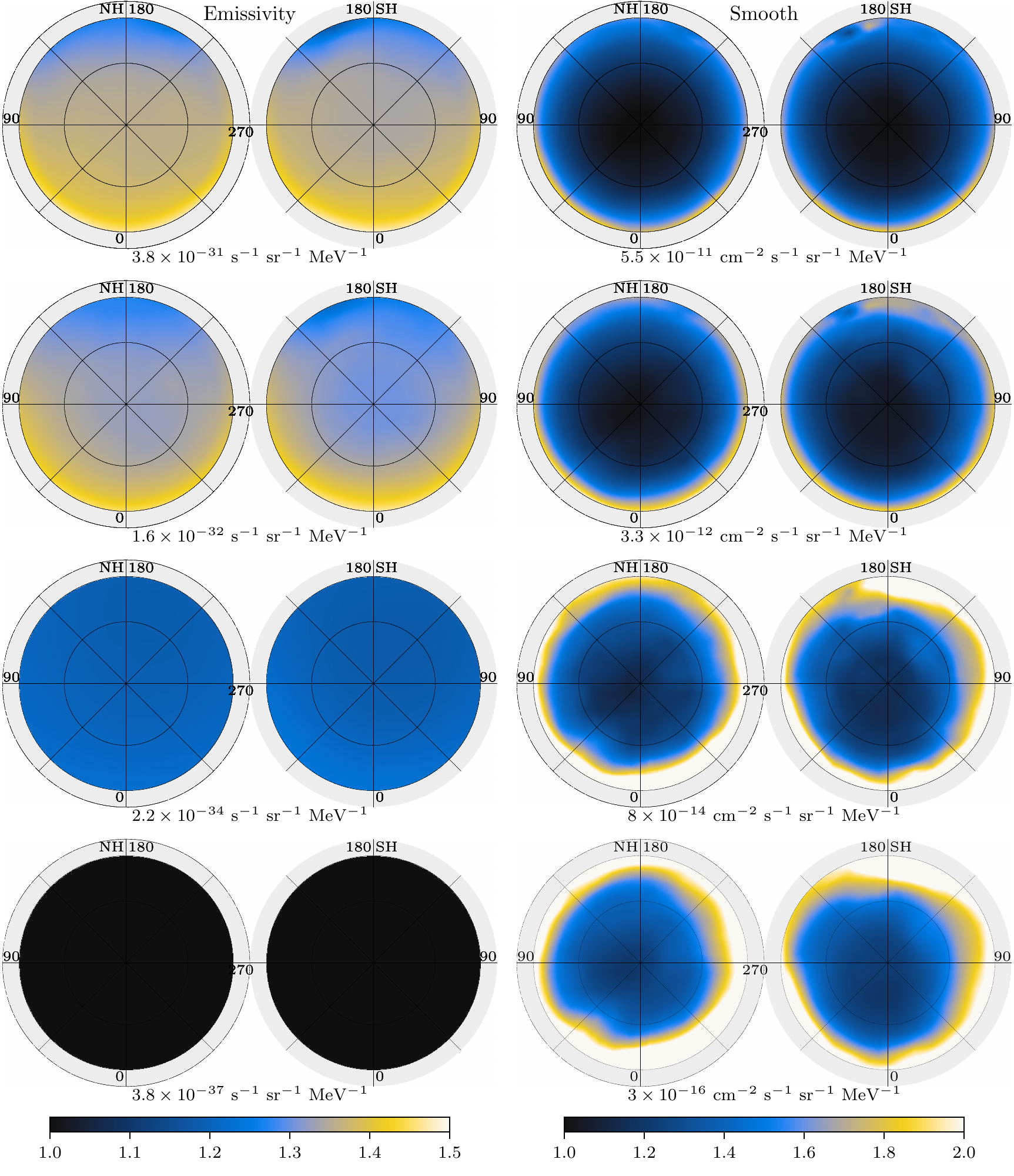}
    \caption{The mean emissivity of the gas (left) and mean intensity of the smooth component (right) for the model results found by analysing the 2000 GALPROP simulations.
      The maps are estimated at the geometric mean of each of the four energy bins, from top to bottom: 1.7 GeV, 5.5 GeV, 24 GeV, and 240 GeV.
      The maps are in Galactic coordinates using an orthographic projection with the NH on left and the SH on right.
      The colour scale is fractionally identical for each column as indicated at the bottom, with the unit scaling given below each map.
      The scale is the same as that used in the \GP\ input truth shown in Fig.~\ref{fig:GALPROP_simulation_truth_maps}. }
    \label{fig:GALPROP_simulation_maps}
\end{figure}

\begin{figure}
    \centering
    \includegraphics{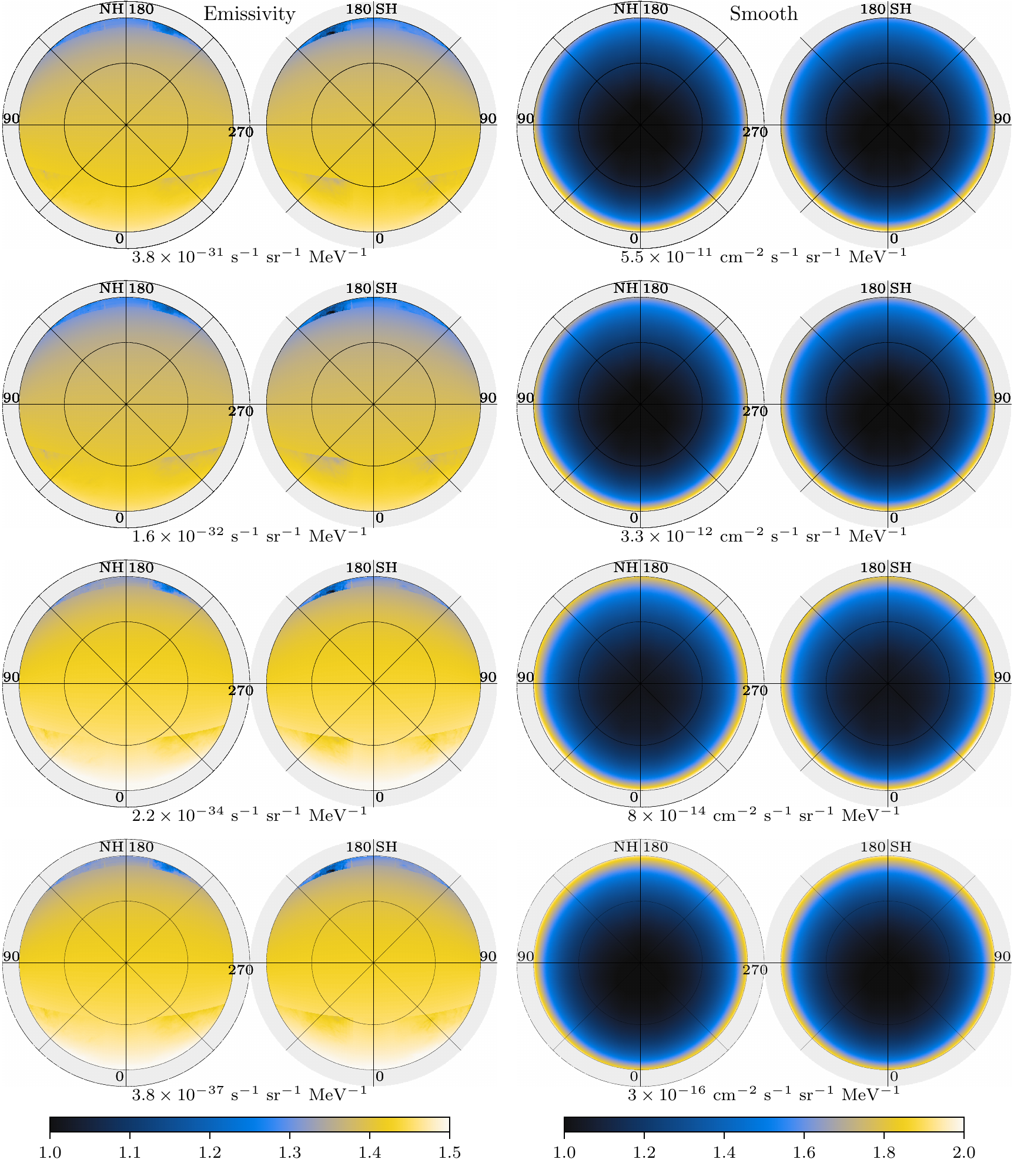}
    \caption{Maps showing the input emissivity of the gas (left) and the intensity of the smooth component (right) \GP\ input to the simulations.
      The maps are estimated at the geometric mean of each of the four energy bins, from top to bottom: 1.7 GeV, 5.5 GeV, 24 GeV, and 240 GeV.
      The maps are in Galactic coordinate using an orthographic projection with the NH on left and the SH on right.
     The colour scale is fractionally identical for each column as indicated at the bottom, with the unit scaling given below each map.}
      The colour scale is the same as that used in the model estimates shown in Fig.~\ref{fig:GALPROP_simulation_maps}.
    \label{fig:GALPROP_simulation_truth_maps}
\end{figure}

Figure~\ref{fig:GALPROP_spectrum} also shows that the analysis recovers the input spectrum reasonably well in the first two energy bins, even though there is a significant curvature in the gas spectrum in the first energy bin.
In the higher two bins, the gas component is under predicted at the level of one to two standard deviation while that of the smooth component is slightly over-predicted.
The corresponding average spatial distributions for each of the four energy bins of the analysis are shown in Fig.~\ref{fig:GALPROP_simulation_maps}.
  %{\bf The mean analysis result spatial distributions for each of the four energy bins the model spectra from Fig~\ref{fig:GALPROP_spectrum} are shown in Fig.~\ref{fig:GALPROP_simulation_maps}.}
For comparison, the input maps are shown in Fig.~\ref{fig:GALPROP_simulation_truth_maps} with the same colour scale.
While the separation of the gas and smooth components is very good in the lowest two energy bins, it is not as successful for the highest energy bin.
The analysis is not able to properly separate out the gas and smooth components in that bin and some of the gas intensity is captured by the smooth component.
The clearest examples of this is the enhancement in the NH towards the inner Galaxy, and in the SH towards the outer Galaxy.
It is clear that due to lack of statistics, the separation of the gas component is inaccurate in the highest energy bin.
The iteration of the input spectrum does not have a significant effect on the spatial results.

\begin{figure}
    \centering
    \includegraphics{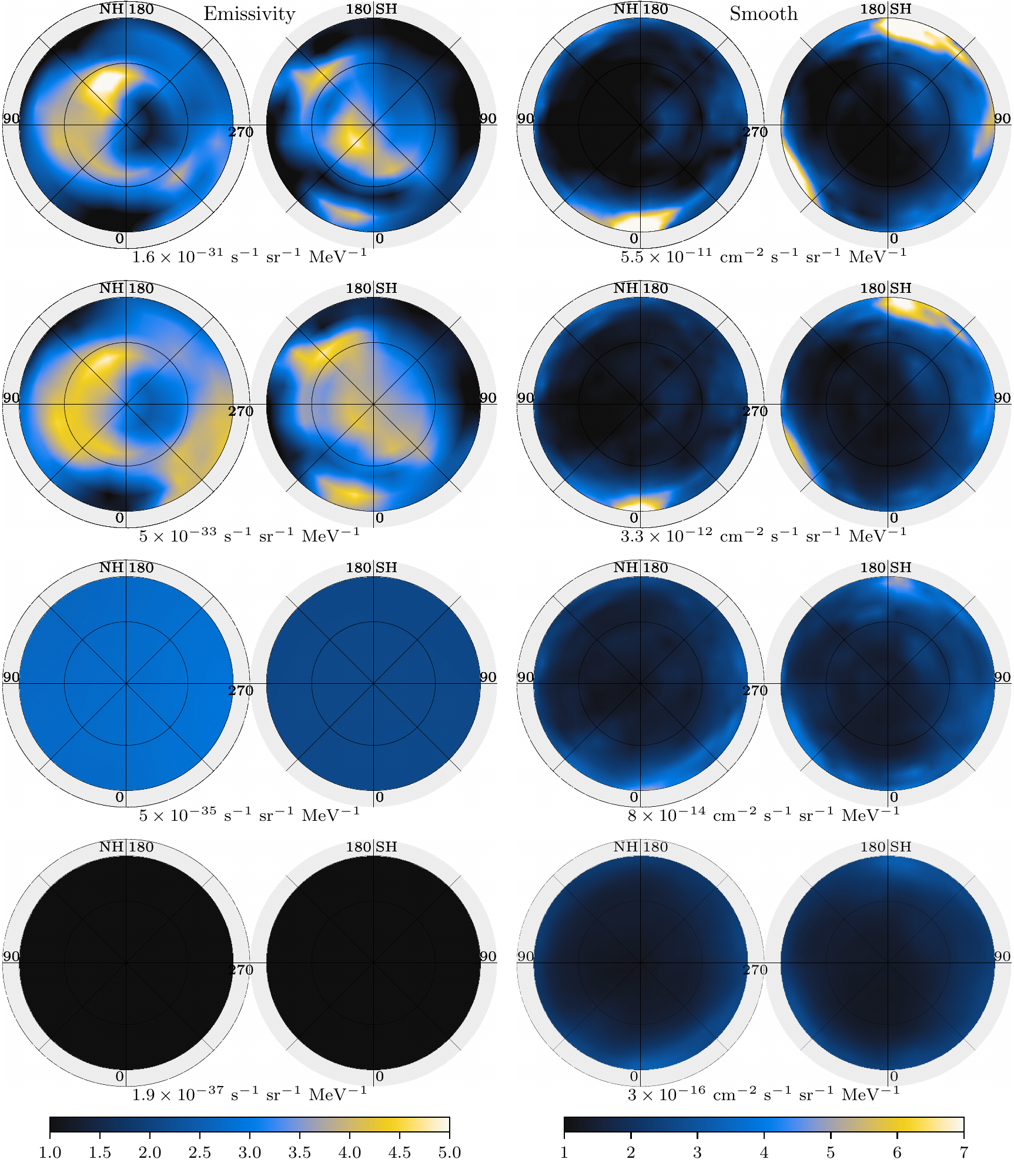}
    \caption{The mean emissivity of the gas (left) and mean intensity of the smooth component (right) for the model results found by analysing the 2000 \GP\ simulations using the $\tau_{353}$ template for the gas.
      The maps are estimated at the geometric mean of each of the four energy bins, from top to bottom: 1.7 GeV, 5.5 GeV, 24 GeV, and 240 GeV.
      The maps are in Galactic coordinates using an orthographic projection with the NH on left and the SH on right.
      The colour scale is fractionally identical for each column as indicated at the bottom, with the unit scale given below each map.
      The colour scale differs significantly from that used in the \GP\ input truth shown in Fig.~\ref{fig:GALPROP_simulation_truth_maps}. }
    \label{fig:GALPROP_simulation_dust_maps}
\end{figure}

\begin{figure}
    \centering
    \includegraphics{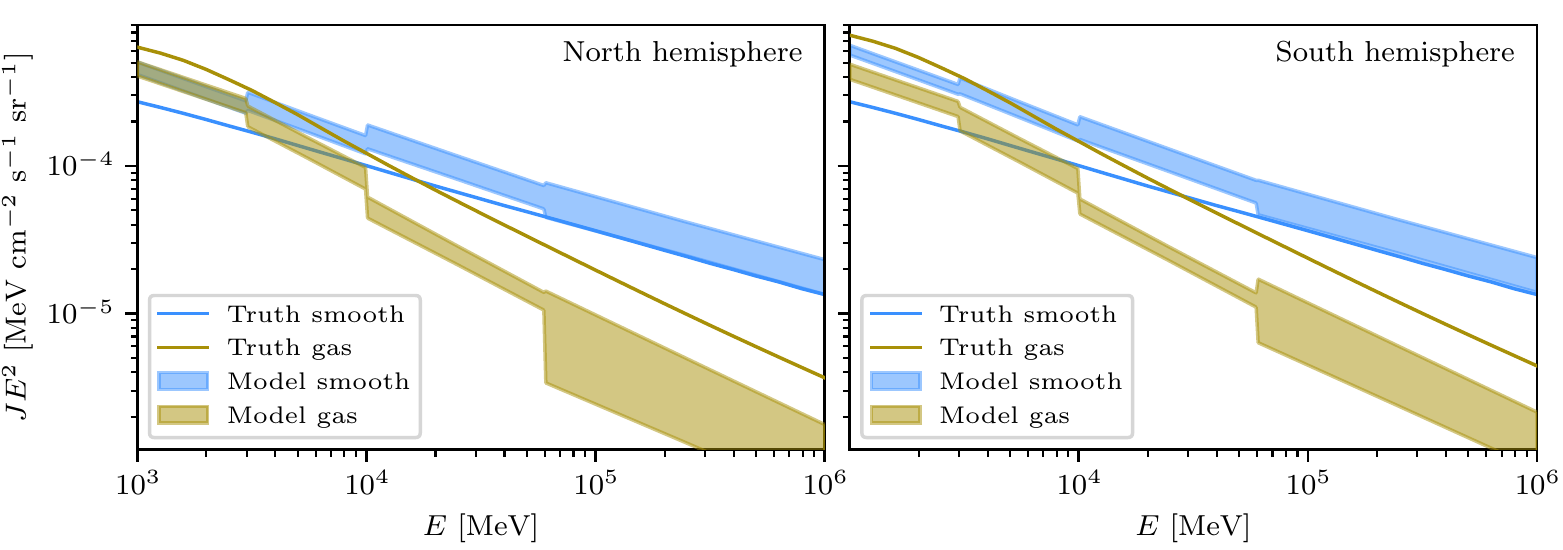}
    \caption{The intensity spectra and 68\% confidence regions of the gas and smooth component after analysing the 2000 simulations that are based on a \GP\ run using the $\tau_{353}$ map as a gas template.
      The green and blue shaded areas show the spectrum of the gas and smooth components, respectively.
      The solid curves show the input truth.  Left panel shows the mean intensity spectrum from the analysis of the NH while the lower panel shows the results using the SH.}
    \label{fig:GALPROP_spectrum_dust}
\end{figure}

The above results of the \GP\ modelling/simulations and analysis have all been performed with the same gas map, which is based on the HI4PI data.
To test the effect of using an incorrect map, the \GP\ results were also analysed using the $\tau_{353}$ map with $\alpha=1.0$.
This has a significant effect on the ability of the analysis to separate out the gas and smooth components, as is evident from the mean intensity of the smooth component and mean emissivity of the gas component shown in  Fig.~\ref{fig:GALPROP_simulation_dust_maps}.
This separation also significantly affects the resulting spectral intensity, as shown in Fig.~\ref{fig:GALPROP_spectrum_dust}.
The analysis results for the gas emissivity and the smooth components intensity are clearly anti-correlated, where one goes up, the other goes down.
This effect significantly exceeds the statistical power of the method and its impact is addressed in Sec.~\ref{subsec:adjust_gas}.

\begin{figure}
    \centering
    \includegraphics{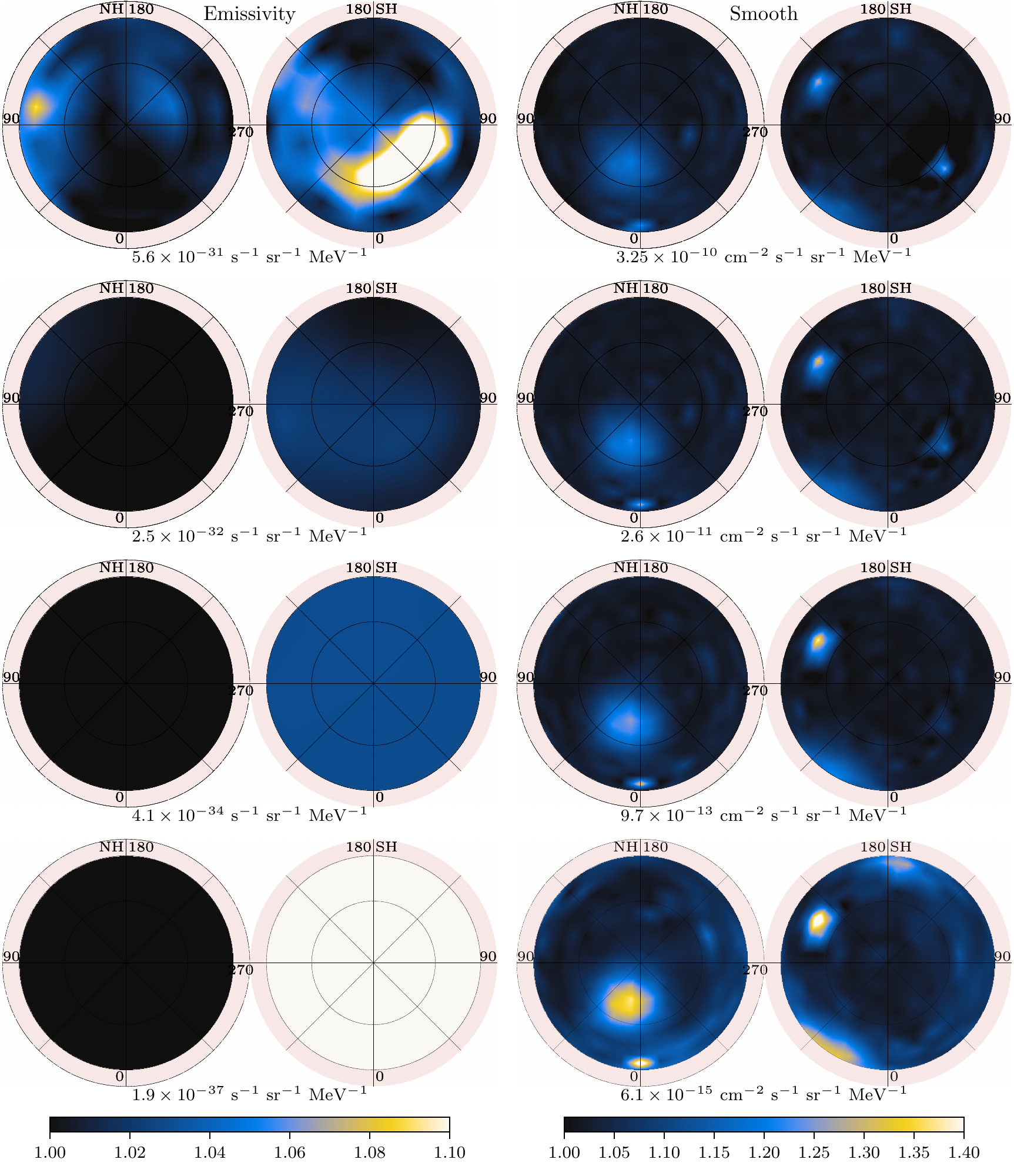}
    \caption{The results from analysis of 2000 simulations based on a model using the 4FGL-DR2 catalogue but analysed using only those sources that are also in the 4FGL.  Left column shows emissivity and right column the intensity of the smooth component.  The colour scale is identical to that used in Fig.~\ref{fig:Bumps_gas} for easy comparison.}
    \label{fig:bumps_4FGL_test}
\end{figure}

To test possible effects that unresolved sources can have on the analysis, simulations are performed utilising the 4FGL-DR2 catalog on top of the three enhanced emission regions described in Section~\ref{subsec:simulations}.
These simulations were then analysed with two different models, one including the full 4FGL-DR2 catalogue to make sure the additional photons would not affect the results and another using only those sources in the 4FGL-DR2 catalogue that are also in the 4FGL catalogue to test the effect of unresolved faint sources.
When using the 4FGL-DR2 catalogue sources as a fixed template, the results are unchanged compared to the results on simulations that did not include any point sources.
For the analysis using only those from the 4FGL catalogue in the fixed source template, the results are shown in Fig.~\ref{fig:bumps_4FGL_test}.
Most of the sources in the 4FGL-DR2 that are not in 4FGL, are faint and do not significantly affect the results, apart from two sources that seem to have been in an enhanced emission state in the time-gap between 4FGL and 4FGL-DR2.
One is in the SH towards $(l,b) \sim (45^\circ,-60^\circ)$ and has a significant effect on the determination of gas emissivity and the smooth component.
The other is in the NH towards $(l,b) \sim (0^\circ, 35^\circ)$ and is visible at a few tens of percent level.
In real data, it is unlikely that such bright sources would not be included in the catalogue, and it is thus safe to conclude that faint unresolved sources should not appear as structured emission in the analysis.

\section{Tabulated Data}
\label{app:data}

Here we provide tabulated data for the spectral quantities plotted in the main text.  
Table~\ref{tab:results_spectrum} provides the data for Fig.~\ref{fig:results_spectrum} and Tables~\ref{tab:results_spectrum_loops} and~\ref{tab:results_spectrum_blobs_split} provide the data for Fig.~\ref{fig:results_spectrum_regions}.  The mean emissivity and IC intensity spectrum of the SA0 model is provided in Table~\ref{tab:SA0_spectrum}.

\begin{table}
  \caption{Data used in Fig.~\ref{fig:results_spectrum}, the average gas emissivity and smooth intensity for $45^\circ < l < 270^\circ$.  The uncertainty provided is the mean of the standard deviation of the per-pixel value for the respective quantities over the region.  Also listed is the intensity of the estimated isotropic component.  Its uncertainty is not estimated.}
  \centering
  \begin{tabular}{c|c|c|c|c|c}
    & \multicolumn{2}{c}{Gas emissivity\tablenotemark{b}} & \multicolumn{2}{c}{Smooth Intensity\tablenotemark{c}} &\\
    $E$\tablenotemark{a} & North & South & North & South & isotropic\tablenotemark{c} \\
    \hline
    1.01  &  14.4 $\pm$ 1.1   & 14.57 $\pm$ 0.96  & 10.50 $\pm$ 0.50  & 9.76 $\pm$ 0.58  & 7.32 \\
    1.32  &  13.2 $\pm$ 1.0   & 12.88 $\pm$ 0.85  &  9.43 $\pm$ 0.45  & 9.00 $\pm$ 0.53  & 6.75 \\
    1.73  &  12.0 $\pm$ 0.9   & 11.38 $\pm$ 0.75  &  8.49 $\pm$ 0.40  & 8.31 $\pm$ 0.49  & 6.23 \\
    2.27  &  11.0 $\pm$ 0.9   & 10.06 $\pm$ 0.66  &  7.63 $\pm$ 0.36  & 7.66 $\pm$ 0.45  & 5.75 \\
    2.97  &  10.1 $\pm$ 0.8   &  8.89 $\pm$ 0.59  &  6.86 $\pm$ 0.33  & 7.07 $\pm$ 0.42  & 5.30 \\
    3.04  &  8.84 $\pm$ 0.61  &  8.95 $\pm$ 0.54  &  7.23 $\pm$ 0.43  & 6.74 $\pm$ 0.51  & 4.74 \\
    4.08  &  7.16 $\pm$ 0.49  &  7.09 $\pm$ 0.43  &  6.56 $\pm$ 0.38  & 6.24 $\pm$ 0.47  & 4.39 \\
    5.48  &  5.80 $\pm$ 0.40  &  5.62 $\pm$ 0.34  &  5.95 $\pm$ 0.35  & 5.78 $\pm$ 0.44  & 4.06 \\
    7.36  &  4.70 $\pm$ 0.33  &  4.45 $\pm$ 0.28  &  5.40 $\pm$ 0.32  & 5.35 $\pm$ 0.40  & 3.76 \\
    9.88  &  3.81 $\pm$ 0.29  &  3.53 $\pm$ 0.23  &  4.90 $\pm$ 0.29  & 4.95 $\pm$ 0.38  & 3.48 \\
    10.2  &  3.79 $\pm$ 0.39  &  3.59 $\pm$ 0.27  &  5.00 $\pm$ 0.43  & 4.73 $\pm$ 0.52  & 3.21 \\
    15.8  &  2.61 $\pm$ 0.25  &  2.47 $\pm$ 0.18  &  4.06 $\pm$ 0.35  & 3.93 $\pm$ 0.43  & 2.67 \\
    24.5  &  1.80 $\pm$ 0.18  &  1.71 $\pm$ 0.14  &  3.30 $\pm$ 0.28  & 3.26 $\pm$ 0.36  & 2.22 \\
    38.0  &  1.24 $\pm$ 0.15  &  1.18 $\pm$ 0.11  &  2.68 $\pm$ 0.23  & 2.71 $\pm$ 0.30  & 1.84 \\
    58.9  &  0.86 $\pm$ 0.12  &  0.81 $\pm$ 0.09  &  2.18 $\pm$ 0.19  & 2.25 $\pm$ 0.25  & 1.53 \\
    61.7  &  1.21 $\pm$ 0.34  &  0.88 $\pm$ 0.21  &  1.86 $\pm$ 0.37  & 1.87 $\pm$ 0.45  & 1.33 \\
    123   &  0.69 $\pm$ 0.18  &  0.52 $\pm$ 0.12  &  1.17 $\pm$ 0.23  & 1.21 $\pm$ 0.29  & 0.858 \\
    245   &  0.39 $\pm$ 0.10  &  0.31 $\pm$ 0.07  &  0.74 $\pm$ 0.14  & 0.78 $\pm$ 0.19  & 0.555 \\
    488   &  0.23 $\pm$ 0.06  &  0.19 $\pm$ 0.05  &  0.47 $\pm$ 0.09  & 0.51 $\pm$ 0.12  & 0.360 \\
    972   &  0.13 $\pm$ 0.04  &  0.11 $\pm$ 0.03  &  0.30 $\pm$ 0.06  & 0.33 $\pm$ 0.08  & 0.233 \\
  \end{tabular}
  \tablenotetext{a}{In units of GeV}
  \tablenotetext{b}{$\epsilon E^2$ in units of $10^{-25}$ MeV sr$^{-1}$ s$^{-1}$}
  \tablenotetext{c}{$J E^2$ in units of $10^{-4}$ MeV sr$^{-1}$ s$^{-1}$ cm$^{-2}$}
  \label{tab:results_spectrum}
\end{table}

\begin{table}
  \caption{Data used in the top two panels of Fig.~\ref{fig:results_spectrum_regions}, the mean of the isotropic subtracted smooth component over the white regions shown in Fig.~\ref{fig:Loops_zoom}.  The uncertainty provided is the mean of the standard deviation of the per-pixel smooth intensity over the region.}
  \centering
  \begin{tabular}{c|c|c|c|c|c}
    $E$\tablenotemark{a}  & Loop I\tablenotemark{b} & Loop IV\tablenotemark{b} & SH $l\sim35$\tablenotemark{b} & North Blob\tablenotemark{b} & South Blob\tablenotemark{b} \\
    \hline
    1.01  &  8.69 $\pm$ 0.72  &  8.62 $\pm$ 0.77  &  4.77 $\pm$ 0.70  &  11.30 $\pm$ 1.06  &  10.91 $\pm$ 0.68 \\
    1.32  &  7.64 $\pm$ 0.64  &  7.57 $\pm$ 0.69  &  4.40 $\pm$ 0.65  &   9.99 $\pm$ 0.96  &  10.06 $\pm$ 0.63 \\
    1.73  &  6.71 $\pm$ 0.58  &  6.65 $\pm$ 0.62  &  4.06 $\pm$ 0.60  &   8.82 $\pm$ 0.86  &   9.29 $\pm$ 0.58 \\
    2.27  &  5.89 $\pm$ 0.52  &  5.83 $\pm$ 0.56  &  3.74 $\pm$ 0.55  &   7.79 $\pm$ 0.78  &   8.57 $\pm$ 0.57 \\
    2.97  &  5.16 $\pm$ 0.47  &  5.11 $\pm$ 0.50  &  3.45 $\pm$ 0.51  &   6.87 $\pm$ 0.71  &   7.91 $\pm$ 0.50 \\
    3.04  &  6.55 $\pm$ 0.63  &  6.02 $\pm$ 0.64  &  4.31 $\pm$ 0.63  &  11.98 $\pm$ 0.95  &   9.35 $\pm$ 0.70 \\
    4.08  &  5.85 $\pm$ 0.57  &  5.37 $\pm$ 0.57  &  3.99 $\pm$ 0.58  &  10.78 $\pm$ 0.86  &   8.66 $\pm$ 0.63 \\
    5.48  &  5.22 $\pm$ 0.51  &  4.79 $\pm$ 0.52  &  3.69 $\pm$ 0.53  &   9.70 $\pm$ 0.78  &   8.02 $\pm$ 0.59 \\
    7.36  &  4.66 $\pm$ 0.47  &  4.27 $\pm$ 0.47  &  3.42 $\pm$ 0.49  &   8.72 $\pm$ 0.71  &   7.43 $\pm$ 0.54 \\
    9.88  &  4.16 $\pm$ 0.43  &  3.80 $\pm$ 0.43  &  3.17 $\pm$ 0.46  &   7.84 $\pm$ 0.66  &   6.88 $\pm$ 0.51 \\
    10.2  &  3.91 $\pm$ 0.60  &  3.49 $\pm$ 0.59  &  3.33 $\pm$ 0.65  &   9.22 $\pm$ 0.99  &   8.46 $\pm$ 0.86 \\
    15.8  &  3.12 $\pm$ 0.48  &  2.78 $\pm$ 0.47  &  2.77 $\pm$ 0.54  &   7.44 $\pm$ 0.79  &   7.02 $\pm$ 0.71 \\
    24.5  &  2.49 $\pm$ 0.39  &  2.21 $\pm$ 0.38  &  2.30 $\pm$ 0.45  &   6.00 $\pm$ 0.65  &   5.83 $\pm$ 0.59 \\
    38.0  &  1.98 $\pm$ 0.32  &  1.76 $\pm$ 0.31  &  1.91 $\pm$ 0.37  &   4.83 $\pm$ 0.53  &   4.84 $\pm$ 0.50 \\
    58.9  &  1.58 $\pm$ 0.27  &  1.40 $\pm$ 0.26  &  1.58 $\pm$ 0.31  &   3.90 $\pm$ 0.44  &   4.02 $\pm$ 0.42 \\
    61.7  &  1.03 $\pm$ 0.47  &  1.05 $\pm$ 0.49  &  1.21 $\pm$ 0.56  &   3.91 $\pm$ 0.96  &   3.99 $\pm$ 0.96 \\
    123   &  0.63 $\pm$ 0.29  &  0.64 $\pm$ 0.30  &  0.78 $\pm$ 0.36  &   2.44 $\pm$ 0.59  &   2.58 $\pm$ 0.61 \\
    245   &  0.38 $\pm$ 0.18  &  0.39 $\pm$ 0.19  &  0.51 $\pm$ 0.23  &   1.52 $\pm$ 0.37  &   1.67 $\pm$ 0.40 \\
    488   &  0.23 $\pm$ 0.12  &  0.24 $\pm$ 0.12  &  0.33 $\pm$ 0.16  &   0.95 $\pm$ 0.24  &   1.08 $\pm$ 0.27 \\
    972   &  0.14 $\pm$ 0.08  &  0.14 $\pm$ 0.08  &  0.21 $\pm$ 0.10  &   0.60 $\pm$ 0.16  &   0.70 $\pm$ 0.19 \\
  \end{tabular}
  \tablenotetext{a}{In units of GeV}
  \tablenotetext{b}{$J E^2$ in units of $10^{-4}$ MeV sr$^{-1}$ s$^{-1}$ cm$^{-2}$}
  \label{tab:results_spectrum_loops}
\end{table}

\begin{table}
  \centering
  \caption{Data used in the bottom panel of Fig.~\ref{fig:results_spectrum_regions}, the mean of the isotropic subtracted smooth component over the white regions shown in Fig.~\ref{fig:Loops_zoom}.  The uncertainty provided is the mean of the standard deviation of the per-pixel smooth intensity over the region.}
  \begin{tabular}{c|c|c|c|c}
    & \multicolumn{2}{c}{North Blob\tablenotemark{b}} & \multicolumn{2}{c}{South Blob\tablenotemark{b}} \\
    $E$\tablenotemark{a}  & High Lat & Low Lat & High Lat & Low Lat \\
    \hline
    1.01  &  12.32 $\pm$ 0.97  &  13.05 $\pm$ 1.25  &  11.30 $\pm$ 0.58  &  15.85 $\pm$ 0.80 \\
    1.32  &  10.91 $\pm$ 0.88  &  11.56 $\pm$ 1.13  &  10.43 $\pm$ 0.53  &  14.63 $\pm$ 0.74 \\
    1.73  &   9.65 $\pm$ 0.79  &  10.24 $\pm$ 1.01  &   9.62 $\pm$ 0.49  &  13.50 $\pm$ 0.68 \\
    2.27  &   8.53 $\pm$ 0.72  &   9.06 $\pm$ 0.92  &   8.88 $\pm$ 0.45  &  12.46 $\pm$ 0.63 \\
    2.97  &   7.53 $\pm$ 0.65  &   8.01 $\pm$ 0.83  &   8.19 $\pm$ 0.43  &  11.49 $\pm$ 0.59 \\
    3.04  &  12.80 $\pm$ 0.85  &  14.24 $\pm$ 1.07  &  10.92 $\pm$ 0.66  &  13.18 $\pm$ 0.84 \\
    4.08  &  11.52 $\pm$ 0.76  &  12.83 $\pm$ 0.96  &  10.11 $\pm$ 0.60  &  12.20 $\pm$ 0.76 \\
    5.48  &  10.36 $\pm$ 0.69  &  11.55 $\pm$ 0.87  &   9.36 $\pm$ 0.55  &  11.30 $\pm$ 0.70 \\
    7.36  &   9.32 $\pm$ 0.63  &  10.40 $\pm$ 0.80  &   8.67 $\pm$ 0.52  &  10.46 $\pm$ 0.65 \\
    9.88  &   8.39 $\pm$ 0.58  &   9.37 $\pm$ 0.74  &   8.03 $\pm$ 0.49  &   9.69 $\pm$ 0.61 \\
    10.2  &   9.95 $\pm$ 0.94  &  11.36 $\pm$ 1.15  &  10.33 $\pm$ 0.89  &  11.56 $\pm$ 1.05 \\
    15.8  &   8.03 $\pm$ 0.76  &   9.18 $\pm$ 0.92  &   8.58 $\pm$ 0.73  &   9.60 $\pm$ 0.86 \\
    24.5  &   6.47 $\pm$ 0.62  &   7.41 $\pm$ 0.75  &   7.12 $\pm$ 0.61  &   7.97 $\pm$ 0.72 \\
    38.0  &   5.22 $\pm$ 0.51  &   5.98 $\pm$ 0.62  &   5.91 $\pm$ 0.52  &   6.62 $\pm$ 0.60 \\
    58.9  &   4.21 $\pm$ 0.43  &   4.83 $\pm$ 0.52  &   4.91 $\pm$ 0.44  &   5.49 $\pm$ 0.51 \\
    61.7  &   4.17 $\pm$ 0.94  &   5.55 $\pm$ 1.28  &   5.72 $\pm$ 1.18  &   4.69 $\pm$ 1.09 \\
    123   &   2.60 $\pm$ 0.57  &   3.47 $\pm$ 0.78  &   3.69 $\pm$ 0.75  &   3.03 $\pm$ 0.69 \\
    245   &   1.63 $\pm$ 0.37  &   2.18 $\pm$ 0.49  &   2.39 $\pm$ 0.50  &   1.96 $\pm$ 0.45 \\
    488   &   1.02 $\pm$ 0.24  &   1.37 $\pm$ 0.32  &   1.55 $\pm$ 0.34  &   1.27 $\pm$ 0.31 \\
    972   &   0.64 $\pm$ 0.16  &   0.86 $\pm$ 0.22  &   1.00 $\pm$ 0.24  &   0.82 $\pm$ 0.21 \\
  \end{tabular}
  \tablenotetext{a}{In units of GeV}
  \tablenotetext{b}{$J E^2$ in units of $10^{-4}$ MeV sr$^{-1}$ s$^{-1}$ cm$^{-2}$}
  \label{tab:results_spectrum_blobs_split}
\end{table}

\begin{table}
  \centering
  \caption{IC intensity and gas emissivity of the SA0 model averaged over $|b|>30^\circ$.}
  \begin{tabular}{c|c|c}
    $E$\tablenotemark{a} & SA0 emissivity\tablenotemark{b} & SA0 IC intensity\tablenotemark{c} \\
    \hline
    1.00 & 14.5   & 2.72 \\
    1.26 & 13.2   & 2.49 \\
    1.58 & 11.8   & 2.27 \\
    2.00 & 10.2   & 2.06 \\
    2.51 & 8.71   & 1.86 \\
    3.16 & 7.38   & 1.69 \\
    3.98 & 6.11   & 1.53 \\
    5.01 & 5.05   & 1.37 \\
    6.31 & 4.12   & 1.24 \\
    7.94 & 3.36   & 1.12 \\
    10.0 & 2.77   & 1.00 \\
    12.6 & 2.28   & 0.903 \\
    15.8 & 1.88   & 0.813 \\
    20.0 & 1.56   & 0.731 \\
    25.1 & 1.30   & 0.661 \\
    31.6 & 1.08   & 0.597 \\
    39.8 & 0.907  & 0.540 \\
    50.1 & 0.760  & 0.490 \\
    63.1 & 0.637  & 0.444 \\
    79.4 & 0.534  & 0.402 \\
    100  & 0.448  & 0.364 \\
    126  & 0.376  & 0.329 \\
    158  & 0.316  & 0.297 \\
    200  & 0.266  & 0.268 \\
    251  & 0.224  & 0.243 \\
    316  & 0.189  & 0.219 \\
    398  & 0.160  & 0.199 \\
    501  & 0.136  & 0.180 \\
    631  & 0.115  & 0.164 \\
    794  & 0.0982 & 0.147 \\
   1000  & 0.0836 & 0.135 \\
  \end{tabular}
  \label{tab:SA0_spectrum}
  \tablenotetext{a}{In units of GeV}
  \tablenotetext{b}{$\epsilon E^2$ in units of $10^{-25}$ MeV sr$^{-1}$ s$^{-1}$}
  \tablenotetext{c}{$J E^2$ in units of $10^{-4}$ MeV sr$^{-1}$ s$^{-1}$ cm$^{-2}$}
\end{table}

\end{document}